\documentclass{jfm}

\usepackage{enumitem}

\usepackage{upgreek}
\usepackage[colorinlistoftodos]{todonotes}

\usepackage{natbib}

\lefttitle{Hao-Chen Liu et al.}
\righttitle{Data-driven DES for turbulent flow}

\title{Data-driven detached-eddy simulations based on explicit algebraic stress expressions for turbulent flows}

\author{Hao-Chen Liu\aff{1,2},
Zifei Yin\aff{3},
Xin-Lei Zhang\aff{1,2}\corresp{\email{zhangxinlei@imech.ac.cn}},
\and
Guowei He\aff{1,2}\corresp{\email{hgw@lnm.imech.ac.cn}}}

\affiliation{\aff{1}The State Key Laboratory of Non-linear Mechanics, Institute of Mechanics, Chinese Academy of Sciences, Beijing 100190, China
  \aff{2}School of Engineering Sciences, University of Chinese Academy of Sciences, Beijing 100049, China
  \aff{3}School of Aeronautics and Astronautics, Shanghai Jiao Tong University, Shanghai 200240, China}

\corresau{Xin-Lei Zhang, zhangxinlei@imech.ac.cn; Guowei He, hgw@lnm.imech.ac.cn}

\begin{document}

%\linenumbers 

\maketitle

\begin{abstract}

This work proposes a data-driven explicit algebraic stress-based detached-eddy simulation (DES) method.
Despite the widespread use of data-driven methods in model development for both Reynolds-averaged Navier-Stokes (RANS) and large-eddy simulations (LES), their applications to DES remain limited.
The challenge mainly lies in the absence of modelled stress data, the requirement for proper length scales in RANS and LES branches, and the maintenance of a reasonable switching behaviour.
The data-driven DES method is constructed based on the algebraic stress equation. 
The control of RANS/LES switching is achieved through the eddy viscosity in the linear part of the modelled stress, under the $\ell^2-\omega$ DES framework. 
Three model coefficients associated with the pressure-strain terms and the LES length scale are represented by a neural network as functions of scalar invariants of velocity gradient.
The neural network is trained using velocity data with the ensemble Kalman method, thereby circumventing the requirement for modelled stress data.
Moreover, the baseline coefficient values are incorporated as additional reference data to ensure reasonable switching behaviour. 
The proposed approach is evaluated on two challenging turbulent flows, i.e., the secondary flow in a square duct and the separated flow over a bump. 
The trained model achieves significant improvements in predicting mean flow statistics compared to the baseline model.
This is attributed to improved predictions of the modelled stress.
The trained model also exhibits reasonable switching behaviour, enlarging the LES region to resolve more turbulent structures. 
Furthermore, the model shows satisfactory generalization capabilities for both cases in similar flow configurations.

\end{abstract}


\section{Introduction}
\label{sec:intro}

Hybrid Reynolds-averaged Navier-Stokes/large eddy simulation (RANS/LES) methods are of practical interest for efficiently predicting massively separated flows in industrial applications \citep{chaouat2017state, heinz2020review}. 
These approaches typically employ the RANS models within the near-wall attached boundary layers and switch to LES in the outer flow, thereby offering a favourable compromise between computational cost and predictive accuracy. 
The detached eddy simulation (DES) \citep{spalart2009detached} is one of the most widely used hybrid approaches, which enables automatic switching from RANS to LES within a unified turbulence model. 
This is achieved by defining the effective length scale as the minimum between the RANS-based integral scale $\ell_{RANS}$ and the LES-based subgrid scale $\ell_{LES}$ (typically proportional to the cell size).
As a result, when the local mesh resolution is sufficiently fine to resolve the associated turbulent structures, the LES mode is activated such that the modelled stress is constrained to be the subgrid-scale stress.

The first DES model, proposed by \citet{spalart1997comments}, enables resolving turbulent structures away from walls by bounding the required wall distance $d$ in the Spalart-Allmaras (SA) turbulence model using the scaled local cell size, \textit{i.e.}, $d^* = \min(d, C_\mathrm{DES}\upDelta)$.
This original DES model primarily suffers from two weaknesses: grid-induced separation (GIS) and log-layer mismatch (LLM) \citep{spalart2009detached}. 
To alleviate the GIS problem, \citet{spalart2006new} developed the Delayed-DES (DDES) model that uses a shielding function to maintain the RANS model behaviour near-walls.
Furthermore, \citet{shur2008hybrid} proposed the improved DDES (IDDES) model, which introduces additional blending functions to enable wall-modelled LES behaviour within boundary layers when the mesh quality allows. 
Meanwhile, the LLM issue is mitigated by redefining the subgrid-scale $\upDelta$ to enable a similar behaviour as the Smagorinsky model on the LES branch.

Alternatively, \citet{reddy2014ddes} proposed the $\ell^2-\omega$ DDES model, offering a relatively simple yet effective framework. 
Specifically, the eddy viscosity in this model is defined using the DDES length scale $\ell_{DDES}$ as $\nu_t = \ell_{DDES}^2 \omega$, which results in an eddy viscosity behaviour that closely resembles the Smagorinsky model in the LES branch. 
As such, the underlying DDES framework can ensure the RANS mode in the near-wall region, avoiding the GIS issue. 
On the other hand, the LLM deficiency can be mitigated by redefining the subgrid-scale $\upDelta$ as the cubic root of the cell volume in the LES region, similar as the IDDES model \citep{shur2008hybrid} but with a simpler definition of $\upDelta$. 
Furthermore, the adaptive version of the $\ell^2-\omega$ DDES model is proposed \citep{yin2015dynamic, yin2016adaptive, bader2022hybrid}, where the DES coefficient $C_\mathrm{DES}$ is dynamically computed based on the Germano identity \citep{lilly1992proposed} or the Vreman kernel \citep{vreman2004eddy}. 
This advancement significantly broadens the applicability of DES across a wide range of flow configurations \citep{yin2021adaptive, yin2022detached, liu2024adaptive}. 

Besides these contributions, numerous extensions and enhancements to the DES method have been explored in the literature \citep{gritskevich2012development,deck2012recent, ashton2013development,le2013zonal,jee2014detached, han2020modification,pont2021new,liu2024improvements}. 
However, most existing DES models are still based on linear eddy viscosity (LEV) RANS models.
These models often struggle to provide reliable predictions in complex flows characterized by strong Reynolds stress anisotropy, such as rotating flows, secondary flows, and flows over curved surfaces \citep{durbin2018some}.

In view of this shortcoming, \citet{liu2024explicit} proposed the explicit algebraic stress-based DDES (EAS-DDES) model.
The modelled stress, denoting the Reynolds stress on the RANS branch and the subgrid stress on the LES branch, is decomposed into a linear and a non-linear part. 
The linear part takes charge of the switching between the RANS and LES branches based on the $\ell^2-\omega$ DDES framework. 
The non-linear part accounts for the modelled stress anisotropy, which is formulated based on the explicit algebraic Reynolds stress (EARS) of \citet{wallin2000explicit}. 
This enables the EAS-DDES model to outperform the LEV-DES models in complex turbulent flows, particularly when near-wall stress anisotropy becomes pronounced\citep{liu2024explicit}.
In addition, the EAS-DDES model can achieve better computational efficiency, stability, and robustness than the Reynolds stress transport-based DES approaches \citep{zhuchkov2016combining, wang2021iddes, liu2021numerical, li2022dynamic} due to the explicit algebraic stress formulation.

The EAS-DDES model is built based on the EARS model of \citet{wallin2000explicit}.
There exist several model coefficients that are determined from canonical turbulent flows.
Specifically, the coefficient values in the pressure-strain rate term of the algebraic stress equation are adopted from \citet{taulbee1992improved, wallin2000explicit}, which is determined based on
homogeneous shear flow \citep{harris1977further}.
For other types of turbulent flows, the model coefficients can be calibrated to provide better predictions \citep{pope2000turbulent}.
Also, the coefficient $C_\mathrm{DES}$ controls the magnitude of the subgrid stress, the standard value of which is determined based on the Smagorinsky constant. 
It has been shown that the adaptive adjustment of $C_\mathrm{DES}$ according to the local turbulent state provides superior performance in various flow cases \citep{yin2015dynamic}. 
Therefore, determining these model coefficients based on local features is promising for further improving the predictive performance of the EAS-DDES model.

Over the past decade, the rapid advancement of data-driven approaches has paved the way for developing predictive turbulence models directly from high-fidelity data \citep{duraisamy2019turbulence, brunton2020machine, duraisamy2021perspectives, sandberg2022machine}. 
Various data-driven methods have been employed to develop RANS turbulence models, including adjoint-based methods \citep{singh2017machine}, decision tree algorithms \citep{matai2019zonal}, sparse regression \citep{schmelzer2020discovery}, gene expression programming \citep{fang2023toward}, random forest \citep{volpiani2024random}, and so on.
Particularly, the ensemble Kalman method~\citep{evensen2009data} has also been applied to train the neural network-based Reynolds stress models \citep{zhang_ensemble-based_2022, zhang2023combining, zhang2023physical}.
This method enables the efficient training of a neural network-based model coupled with the RANS solver, thereby avoiding inconsistencies between the training and prediction environments~\citep{duraisamy2021perspectives}. 
Moreover, the data requirement is significantly relaxed, as only measurable flow quantities, such as sparse measurements of mean velocities, are needed, rather than full-field Reynolds stress. 

Besides the applications in RANS models, data-driven methods are also emerging for the LES in building subgrid stress models and wall models \citep{duraisamy2021perspectives}. 
Early works mainly focus on subgrid stress modelling of canonical turbulent flows, such as the isotropic turbulence \citep{sirignano2020dpm} and channel flow \citep{park2021toward, xu2023artificial}. 
Further, \citet{sirignano2023deep} trained a subgrid stress model for bluff-body separation flows using adjoint-based methods,
showing significant improvement in predictive accuracy over traditional models.
The data-driven method has also been applied to build neural network-based wall models \citep{zhou2021wall, zhou2025wall}.
Particularly, \citet{lozano2023machine} proposed a neural network-based building-block-flow wall model, which has been applied to two realistic aircraft configurations.
More recently, \citet{zhang2025knowledge} proposed a knowledge-integrated additive learning approach for learning LES wall models, the predictive capability of which has been demonstrated in channel flows, separated flows over periodic hills, and the 2-D Gaussian bump.

Despite the widespread use of data-driven methods in the development of both RANS and LES models, their application to the DES remains limited.
The primary challenges of the data-driven DES stem from three key aspects.
The first is the absence of a benchmark database for modelled stress. In data-driven RANS modelling, the target data of Reynolds stress is the second-order moments of the velocity fluctuations. In data-driven LES modelling, the target data of subgrid stress is also accessible from the filtered DNS.
However, in a DES simulation, the switching process from RANS to LES trades the modelled and the resolved stresses.
No instantaneous, explicit ``optimal'' switching from RANS to LES can be derived. As such, the local proportion of the total stress to be resolved and modelled is unknown and cannot be obtained from DNS or experimental measurements.

Given that DES is intrinsically an automatic length scale formulation, the second challenge is to ensure the proper length scales for the model to choose from.
Specifically, a too low RANS length scale in the region away from the wall may cause the LES subgrid viscosity to be wrongly suppressed when resolving eddies.
Correspondingly, a spuriously low LES length scale in the near-wall region would also cause modelled stress depletion, which may lead to GIS or LLM issues \citep{spalart2009detached}.
It is a new constraint for data-driven modelling because no previous practice requires consideration of how the model would perform when an ``LES'' solution is fed to a data-driven RANS model, and vice versa.

The third challenge lies in maintaining a seamless and stable switching between RANS and LES branches, particularly in flows for which the model has not been trained.
The RANS-to-LES interface continuously evolves at each time step during the simulation. 
Ideally, it should become stabilized within a certain range of wall distance, regardless of the initial condition.
However, as the data-driven model is trained statistically on certain flow types, its generalizability in maintaining proper interface location is difficult to guarantee.
Drifting away from the ideal interface location during the unsteady time-marching can lead to LLM or even become a pure RANS or coarse-grid LES.
It remains an open issue in the hybrid RANS/LES modelling community to achieve seamless switching between RANS and LES.
And it is even more challenging in data-driven DES modelling, given the non-linear nature and the indirect tuning of the coefficients of neural networks.

The present study aims to propose a framework for building a data-driven DES model based on the algebraic stress equation. 
To the authors’ knowledge, this is the first such attempt in data-driven turbulence modelling for the DES.
A physically reasonable and mathematically consistent expression of the modelled stress in RANS and LES branches is constructed through the algebraic stress approach under the weak-equilibrium assumption.
The switching between the RANS and LES branches is achieved through the eddy viscosity, or equivalently, turbulent kinetic energy (TKE). 
The neural network is used to represent the functional mapping from the local flow features to the model coefficients. 
The neural network is trained using the ensemble Kalman method, which enables model training with mean velocity data from experiments or high-fidelity simulations, thereby avoiding the requirement for modelled stress data.
The reasonable length scales and switching behaviours are enforced by augmenting the training data with the baseline values of the model coefficients, which can ensure similar behaviours to the baseline model.

The rest of this paper is organized as follows. 
The general expression of the modelled stress in DES based on the algebraic stress modelling approach is established in \S \ref{sec:modelling}. 
The data-driven closure of the model coefficients, including the neural network-based model representation and the ensemble Kalman method for model training, is introduced in \S \ref{sec:DD}. 
The capability of the present data-driven explicit algebraic stress DDES (DD-EAS-DDES) approach in predicting complex turbulent flows, specifically the secondary flow in a square duct and the separated flow over a bump, is demonstrated in \S \ref{sec:duct} and \S \ref{sec:bump}, respectively. 
The generalizability of the present model is also illustrated. 
The physical consistency and training efficiency of the present approach are further discussed in \S \ref{sec:discussion}. 
Finally, \S \ref{sec:conclusion} concludes the entire work.

\section{Modelling framework}
\label{sec:modelling}

\subsection{Governing equations}

The governing equations for the EAS-DDES model \citep{liu2024explicit} read as 
\begin{equation}
\label{toteq}
\begin{aligned}
    &\bnabla \cdot \boldsymbol{ \bar{u}} = 0, \\
    &\frac{\mathrm{D} \boldsymbol{ \bar{u}}}{\mathrm{D} t} = -\bnabla \bar{p} + \bnabla \cdot\left( \nu  \bnabla \bar{u}\right) - \bnabla \cdot \boldsymbol{\tau}, \\
    &\frac{\mathrm{D} k}{\mathrm{D} t} =  - \boldsymbol{\tau} : \boldsymbol{\bar{S}} -C_\mu  k \omega+\bnabla \cdot\left[ \left(\nu + \frac{C_\mu^* \sigma_k}{C_\mu}  \frac{k}{\omega} \right) \bnabla k\right], \\
    &\frac{\mathrm{D} \omega}{\mathrm{D} t} =  \frac{2C_{\omega 1}C_\mu^*}{C_\mu} |\bar{S}|^2 - C_{\omega 1}\omega \boldsymbol{a^*} : \boldsymbol{\bar{S}} - C_{\omega 2}  \omega^2+\bnabla \cdot\left[ \left(\nu +  \frac{C_\mu^* \sigma_\omega}{C_\mu}  \frac{k}{\omega} \right) \bnabla \omega\right],
\end{aligned}
\end{equation}
 where $p$ denotes the pressure normalized by the constant flow density, $\boldsymbol{u}$ is the velocity, $\nu$ is the kinetic viscosity, $k$ is the TKE, $\omega$ is the turbulence frequency, $C_\mu$ is the standard eddy viscosity coefficient, $C_\mu^*$ represents the effective $C_\mu$, and $\boldsymbol{\bar{S}}$ denotes the strain rate tensor as 
\begin{equation}
\label{Sbar}
       \boldsymbol{\bar{S}} \equiv \tfrac{1}{2} \left[ \bnabla\boldsymbol{ \bar{u}} + (\bnabla \boldsymbol{ \bar{u}})^\mathrm{T} \right] \text{.}
\end{equation}
Here, the RANS ensemble-average and the LES spatial-filtering are collectively denoted by the overline $\bar{(\cdot)}$. The standard values of the model constants are adopted as $C_\mu = 0.09$, $\sigma_k = \sigma_\omega = 0.5$, $C_{\omega 1} = 5/9$, $C_{\omega 2} = 3/40$.

The modelled stress $\boldsymbol{\tau}$ is expressed as \citep{liu2024explicit}
\begin{equation}
\label{tauij}
    \boldsymbol{\tau} \equiv \overline{\boldsymbol{uu}} - \bar{\boldsymbol{u}} \bar{\boldsymbol{u}} =  k \boldsymbol{a} + \tfrac{2}{3}k \boldsymbol{I} =  -2\nu_t \boldsymbol{\bar{S}}  + k \boldsymbol{a^*} + \tfrac{2}{3}k \boldsymbol{I}.  
\end{equation}
In the EAS-DDES model, the linear part~$(-2\nu_t\bar{S})$ directly controls the switching between the RANS and LES branches through the eddy viscosity $\nu_t$, and the non-linear part~$\boldsymbol{a^*}$ reacts to that, adding the extra anisotropy to the modelled stress.
In the following, the modelling of the linear and non-linear parts of the turbulent stress is illustrated, respectively.

\subsection{Modelling of the linear part in the DDES manner\label{sec:modelDDES}}

The linear part of the turbulent stress is modelled by following the $\ell^2-\omega$ DDES framework \citep{reddy2014ddes}.
Specifically, the eddy viscosity $\nu_t$ in the linear part is calculated as
\begin{equation}
\label{DESnut}
    \nu_t = \ell_{DDES}^2 \omega \text{.}  
\end{equation}
The DDES length scale $\ell_{DDES}$ calculated from the RANS and LES length scales ($\ell_{RANS}$ and $\ell_{LES}$) as
\begin{equation}
    \ell_{DDES} = \ell_{RANS} - f_d \max (0, \ell_{RANS} - \ell_{LES}),
\end{equation}
where
\begin{equation}
\label{lengths}
\begin{gathered}
    \ell_{RANS} = \sqrt{\frac{C_\mu^*}{C_\mu}} \frac{\sqrt{k}}{\omega}, \quad \ell_{LES} = C_\mathrm{DES} \upDelta,\\
\upDelta=f_d V^{1 / 3}+\left(1-f_d\right) h_{\max } .
\end{gathered}
\end{equation}
Here $V$ is the local cell volume, and $h_{max}$ is the local maximum cell spacing in the three directions.
The shielding function $f_d$ is to ensure the near-wall RANS region, which is formulated as
\begin{equation}
\label{fd}
    f_d = 1 - \tanh [(8r_d)^3], \quad r_d = \frac{(C_\mu^*/ C_\mu)k/\omega + \nu}{\kappa^2 d_w^2 \sqrt{\bnabla\boldsymbol{ \bar{u}} : \bnabla\boldsymbol{ \bar{u}}}}
\end{equation}
with $\kappa$ being the Von Kármán constant, and $d_w$ the wall distance.

\subsection{General expression of the non-linear part}

As for the non-linear extra anisotropy $\boldsymbol a^*$, the most general expression is adopted as~\citep{pope1975more}
\begin{equation}
\label{afull}
\boldsymbol{a^*} = \sum_{i=2}^{10} \beta_i \boldsymbol{T}_i, 
\end{equation}
which is derived under the effective-viscosity hypothesis and the Cayley-Hamilton theorem. 
Here $\boldsymbol{T}_i$ are the tensor basis as 
\begin{equation}
\label{tensorbasis}
\begin{aligned}
&\boldsymbol{T}_2 = \boldsymbol{S}^2 - \tfrac{1}{3} \theta_1 \boldsymbol{I} 
, \quad \boldsymbol{T}_3 = \boldsymbol{\Omega}^2 - \tfrac{1}{3} \theta_2 \boldsymbol{I} , \quad \boldsymbol{T}_4 = \boldsymbol{S}\boldsymbol{\Omega} - \boldsymbol{\Omega}\boldsymbol{S} , \\
&\boldsymbol{T}_5 = \boldsymbol{S}^2\boldsymbol{\Omega} - \boldsymbol{\Omega}\boldsymbol{S}^2 , \quad 
\boldsymbol{T}_6 = \boldsymbol{S}\boldsymbol{\Omega}^2 + \boldsymbol{\Omega}^2\boldsymbol{S}  - \tfrac{2}{3} \theta_4 \boldsymbol{I} - \theta_2 \boldsymbol{S}, \\
& \boldsymbol{T}_7 = \boldsymbol{S}^2\boldsymbol{\Omega}^2 + \boldsymbol{\Omega}^2\boldsymbol{S}^2  - \tfrac{2}{3} \theta_5 \boldsymbol{I} - \theta_4 \boldsymbol{S} , \quad 
\boldsymbol{T}_8 = \boldsymbol{S}\boldsymbol{\Omega}\boldsymbol{S}^2 - \boldsymbol{S}^2\boldsymbol{\Omega}\boldsymbol{S},  \\
& \boldsymbol{T}_9 = \boldsymbol{\Omega}\boldsymbol{S}\boldsymbol{\Omega}^2 - \boldsymbol{\Omega}^2\boldsymbol{S}\boldsymbol{\Omega} , \quad \boldsymbol{T}_{10} = \boldsymbol{\Omega}\boldsymbol{S}^2\boldsymbol{\Omega}^2 - \boldsymbol{\Omega}^2\boldsymbol{S}^2\boldsymbol{\Omega} ,
\end{aligned}
\end{equation}
with $\boldsymbol I$ being the identity matrix.
The tensor coefficients $\beta_i$ depend on the five invariants as 
\begin{equation}
\label{theta}
   \theta_1 = {\rm tr} (\boldsymbol{S}^2), \quad 
   \theta_2 = {\rm tr} (\boldsymbol{\Omega}^2)  , \quad 
   \theta_3 = {\rm tr} ( \boldsymbol{S}^3 ) , \quad 
   \theta_4 = {\rm tr} ( \boldsymbol{S \Omega}^2 )  , \quad 
   \theta_5 = {\rm tr} ( \boldsymbol{S}^2 \boldsymbol{\Omega}^2 ) \text{.}
\end{equation}
The normalized strain and rotation rates are defined as
\begin{equation}
\label{S_Omega}
    \boldsymbol{S} \equiv \frac{\bnabla\boldsymbol{ \bar{u}} + (\bnabla \boldsymbol{ \bar{u}})^\mathrm{T}}{2C_\mu \omega} , \quad \boldsymbol{\Omega} \equiv -\frac{\bnabla \boldsymbol{\bar{u}} - (\bnabla \boldsymbol{\bar{u}})^\mathrm{T}}{2C_\mu \omega} \text{.}
\end{equation}

Note that the tensor bases $\boldsymbol{T}_6$ and $\boldsymbol{T}_7$ differ from the original expressions of \citet{pope1975more} in the existence of the last terms.
This is due to the extraction of the linear components, which are absorbed into the linear part of the modelled stress. 
Accordingly, $C_\mu^*$ is expressed as 
\begin{equation}
\label{Cmustar}
    C_\mu^{*} = -\tfrac{1}{2}(\beta_1 + \theta_2\beta_6 + \theta_4 \beta_7)
\end{equation}
where $\beta_1$ represents the coefficient of the first tensor basis $\boldsymbol{T}_1 = \boldsymbol{S}$ in the general expression of $\boldsymbol{a}$.

\subsection{Closure through the algebraic stress equation}

To close the tensor basis coefficients $\beta_i$, the algebraic stress modelling framework is adopted under the weak-equilibrium assumption.
The transport of the modelled stress $\boldsymbol{\tau}$ can be considered to be due to transports in $k$ and modelled stress anisotropy $\boldsymbol{a}$. The former is retained, while the latter is neglected \citep{pope2000turbulent}, 
resulting in an implicit algebraic equation of $\boldsymbol{a}$ as
\begin{equation}
\label{ARSM}
(\boldsymbol{a} + \tfrac{2}{3}\boldsymbol{I})(\mathcal{P}-\varepsilon)= \boldsymbol{P}-\boldsymbol{E}+\boldsymbol{\Pi} \text{,}
\end{equation}
where the production tensor $\boldsymbol{P}$ is in closed form as
\begin{equation}
    \label{pij}
    \boldsymbol{P} \equiv - \boldsymbol{\tau} \bnabla\boldsymbol{ \bar{u}} - (\bnabla\boldsymbol{ \bar{u}})^\mathrm{T} \boldsymbol{\tau} 
    = \varepsilon (-\tfrac{4}{3}\boldsymbol{S} - \boldsymbol{aS} - \boldsymbol{Sa}  + \boldsymbol{a \Omega} - \boldsymbol{\Omega a}) \text{,}
\end{equation}
and $\mathcal{P} \equiv {\rm tr}(\boldsymbol{P})/2$. 
For the dissipation tensor $\boldsymbol{E}$, lumping the dissipation anisotropy into the pressure-strain rate tensor $\boldsymbol{\Pi}$ results in~\citep{pope2000turbulent}
\begin{equation}
    \label{eij}
    \boldsymbol{E} = \tfrac{2}{3} \varepsilon \boldsymbol{I} \text{.}
\end{equation}

The pressure-strain rate tensor $\boldsymbol{\Pi}$ can be decomposed into a slow and a rapid component, i.e., $\boldsymbol{\Pi}^s$ and $\boldsymbol{\Pi}^r$, respectively. 
The Rotta's model \citep{rotta1951statistische} is used for $\boldsymbol{\Pi}^s$ as
\begin{equation}
\label{rotta}
    \boldsymbol{\Pi}^s = -C_1 \varepsilon \boldsymbol{a} \text{,}
\end{equation}
representing a linear return to isotropy of the Reynolds stress. 
The Rotta constant $C_1$ determines the rate of return to isotropy. 
As for the rapid pressure-strain rate $\boldsymbol{\Pi}^r$, there exist several models \citep{pope2000turbulent}. 
Here we follow \citet{taulbee1992improved} and \citet{wallin2000explicit} to adopt the general linear model of \citet{launder1975progress} for $\boldsymbol{\Pi}^{r}$ as
\begin{equation}
\label{LRR}
    \boldsymbol{\Pi}^{r}=-\frac{C_2+8}{11}\left(\boldsymbol{P}-\frac{2}{3} \mathcal{P} \boldsymbol{I}\right)-\frac{60 C_2-4}{55} \varepsilon\boldsymbol{S}-\frac{8 C_2-2}{11}\left(\boldsymbol{D}-\frac{2}{3} \mathcal{P} \boldsymbol{I}\right) \text{,}
\end{equation}
where
\begin{equation}
\label{D}
    \boldsymbol{D} \equiv - \boldsymbol{\tau} (\bnabla\boldsymbol{ \bar{u}})^\mathrm{T} - (\bnabla\boldsymbol{ \bar{u}}) \boldsymbol{\tau} 
    = \varepsilon (-\tfrac{4}{3}\boldsymbol{S} - \boldsymbol{aS} - \boldsymbol{Sa}  - \boldsymbol{a \Omega} + \boldsymbol{\Omega a}) \text{.}
\end{equation}

Inserting equation \eqref{pij}-\eqref{D} into \eqref{ARSM}, the algebraic equation for the modelled stress anisotropy tensor $\boldsymbol{a}$ is derived as
\begin{equation}
\label{ARSM1}
\left(C_1 - 1 + \frac{\mathcal{P}}{\varepsilon} \right) \boldsymbol{a}=-\frac{8}{15} \boldsymbol{S}
+\frac{7C_2 +1}{11}(\boldsymbol{a} \boldsymbol{\Omega}-\boldsymbol{\Omega} \boldsymbol{a})-\frac{5-9C_2}{11}\left[\boldsymbol{a S}+\boldsymbol{S a}-\tfrac{2}{3} \operatorname{tr} (\boldsymbol{a} \boldsymbol{S}) \boldsymbol{I} \right] \text{,}
\end{equation}
where
\begin{equation}
\label{pe}
\mathcal{P}/\varepsilon \equiv -\operatorname{tr} (\boldsymbol{a} \boldsymbol{S}) \text{.}
\end{equation}
The above algebraic stress equation is a non-linear equation of $\boldsymbol a$ due to equation \eqref{pe}. 
\citet{wallin2000explicit} obtained the approximated solution by inserting the general form of $\boldsymbol{a}$ as equation \eqref{afull} into \eqref{ARSM1}. 
The resulting linear system of $\beta_i$ can then be formulated as functions of $\mathcal{P}/\varepsilon, C_1, C_2$ and $\theta_j$ by assuming that $\mathcal{P}/\varepsilon$ is already determined. Then $\mathcal{P}/\varepsilon$ is solved by inserting the solution of $\boldsymbol{a}$ into \eqref{pe}. 
In doing so, the functions of $\beta_i (C_1, C_2; \theta_j)$ can be obtained. 
The full expressions of the tensor coefficients~$\beta$ are provided in Appendix \ref{sec:appA}. 
The neglect of the advection and diffusion of $\boldsymbol{a}$ under the weak equilibrium assumption may cause problems in flows where the production-to-dissipation ratio is small, leading to overestimated $C_\mu^*$. 
Given this, an effective diffusion term is added as a correction in such cases~\citep{wallin2000explicit}.
The details are also provided in Appendix~\ref{sec:appA}.
As such, the present general EAS-DDES model is obtained, provided with the pressure-strain rate coefficients $C_1, C_2$ and the DES coefficient $C_\mathrm{DES}$.

Note that the algebraic equation \eqref{ARSM1} of $\boldsymbol{a}$ is consistent for both the Reynolds stress and the subgrid stress. 
Although the extra correlation terms $\overline{\overline{\boldsymbol{S}} p'}$ exist in the pressure-strain rate tensor due to the filtering process, $\boldsymbol{\Pi}$ can be modelled in a similar way for both the Reynolds stress and the subgrid stress as equations \eqref{rotta} and \eqref{LRR} \citep{marstorp2009explicit}.
Hence, equation \eqref{ARSM1} and its solution $\beta_i(C_1, C_2; \theta_i)$ are mathematically consistent in both the RANS and eddy-resolving regions, for the hybrid RANS/LES methods.

The present EAS-DDES framework is based on the weak-equilibrium assumption \citep{rodi1972prediction}, where the convection and diffusion of the anisotropy $\boldsymbol{a}$ are neglected.
Hence, the variation of the Reynolds stress is attributed solely to the modelled kinetic energy, $k$, i.e., 
\begin{equation}
    \frac{\mathrm{D} \boldsymbol{\tau}}{\mathrm{D} t} - \boldsymbol{T} \approx \frac{\boldsymbol{\tau}}{k} \left(\frac{\mathrm{D} k}{\mathrm{D} t} - T_k  \right)
\end{equation}
where $\boldsymbol{T}$ is the diffusion of modelled stress and $T_k = \frac{1}{2}\operatorname{tr} (\boldsymbol{T})$ is the diffusion of $k$. 
In the DES method, the switching of $k$ from the TKE in the RANS branch to subgrid kinetic energy is achieved by the suppression of $\nu_t$, 
which consequently affects the production term in the transport equation of $k$. 
Therefore, the modelled stress can be naturally switched from the Reynolds stress to the subgrid stress under the weak-equilibrium assumption, leading to a consistent form of the modelled stress anisotropy $\boldsymbol{a}$ in RANS and LES branches.

Admittedly, an algebraic stress model under the weak-equilibrium assumption cannot fully represent transport effects like a full differential Reynolds stress model. 
However, it offers a significant improvement over LEV models by capturing anisotropy effects. The algebraic closure of the modelled stress anisotropy $\boldsymbol{a}$ provides a reasonable compromise between the computational cost and physical accuracy for many practically relevant flows \citep{gatski2000nonlinear, durbin2018some}.

For the baseline EAS-DDES model \citep{liu2024explicit}, the coefficients $C_1 = 1.8$ and $C_2 = 5/9$ are set by following \citet{wallin2000explicit}, and the coefficient $C_\mathrm{DES} = 0.12(C_{\mu}^*/C_{\mu})^{-1/4}$ is set to ensure the same subgrid stress intensity as the Smagorinsky model with the Smagorinsky constant $C_S = 0.2$. 
The last term in equation \eqref{ARSM1} vanishes at $C_2 = 5/9$, and the solution of $\beta_i$ is largely simplified due to $\beta_{2,5,7,8,10} = 0$.
In the next section, we introduce the data-driven method to learn the neural network-based closure of these coefficients.

\section{Data-driven closure of the model coefficients}
\label{sec:DD}

As described in \S\ref{sec:modelling}, the general EAS-DDES model is closed with the model coefficients $C_1$, $C_2$, and $C_\mathrm{DES}$ determined. 
Although the baseline model \citep{liu2024explicit} provides a set of coefficients, it is of significant interest to optimize the coefficient set for further predictive improvement. 
For the pressure-strain rate coefficients $C_1$ and $C_2$, various combinations have been employed in previous studies, and the optimal set of coefficients is found to differ across various cases \citep{pope2000turbulent}. 
In addition, the modelled stress anisotropies at the RANS integral scale and the LES subgrid scale are inherently different, requiring different sets of $C_1$ and $C_2$ on the RANS and LES branches. 
This is evident in the explicit algebraic subgrid stress model of \citet{marstorp2009explicit}, which employs a locally adaptive coefficient $C_1$. It is also expected that the adaptive coefficients based on local flow features can partly compensate for the errors arising from the weak-equilibrium assumption by capturing some neglected advection and diffusion effects in the Reynolds stress anisotropy. 
An example is the correction proposed by \citet{wallin2000explicit} that adjusts the model coefficient $C_1$ in regions with low $\mathcal{P}/\varepsilon$, as provided in Appendix~\ref{sec:appA}.

The DES constant $C_\mathrm{DES}$ is also not universal.
Different values have been used in various versions of DES models, e.g., 0.65 for Spalart-Allmaras-based DDES \citep{spalart2006new}, 0.78 and 0.61 for $k-\omega$ SST-based DDES \citep{gritskevich2012development}, and 0.12 for $\ell^2-\omega$ DDES \citep{reddy2014ddes}. 
Particularly, \citet{yin2015dynamic} proposed a dynamic version for $C_\mathrm{DES}$, which exhibits superior performance in various cases, and is necessary for a correct model behaviour in a transitional flow \citep{yin2021adaptive}. 
Given these findings, the present study aims to offer a data-driven DES method to construct the adaptive model coefficients $C_1, C_2$, and $C_\mathrm{DES}$ based on local flow features. 

\subsection{Neural network-based representation of model coefficients}
\label{sec:nnmodel}

The present work employs a fully connected neural network to represent the mapping from local field variables to the model coefficients. 
Selecting appropriate input features is crucial for obtaining an accurate and generalizable model. 
Various features have been used in data-driven turbulence models \citep{duraisamy2021perspectives}. 
In the present work, the five invariants $\theta_i$ (as defined in equation \eqref{theta}) are used as input features for the neural network. 
This is in light of the fact that the tensor basis coefficients $\beta_i$ can generally be expressed as functions of these invariants \citep{pope1975more}. 
Moreover, the same features have been used in the neural network-based closure of the Reynolds stress \citep{ling2016reynolds} and subgrid stress \citep{bose2024invariance}.

The input features are scaled into the range of $[-1,1]$ by $\theta_i^* = \theta_i/(| \theta_i | +1 )$ to facilitate the training convergence. This approach falls into the category of input normalization based on local quantities, $\theta_i^* = \hat{\theta}_i/(| \hat{\theta}_i | + \theta_0 )$, where $\hat{\theta}$ is the dimensional scalar invariants, calculated from the dimensional strain and rotation rates, and $\theta_0$ is the local normalization factor defined from the turbulent time scale $1/ (C_\mu \omega)$.
This has been widely adopted in data-driven turbulence modelling \citep{wang2017physics, wu2018physics}. 
In the present study, the dimensional scalar invariants $\hat{\theta}_i$ are normalized based on $C_\mu \omega$ to obtain $\theta_i$, leading to the equivalent expression of $\theta_i^* = \theta_i/(| \theta_i | +1 )$.
This approach has been shown to yield satisfactory training performance~\citep{ling2015evaluation,wu2025framework}.

Moreover, to improve numerical stability, the local coefficients $C_1, C_2$, and $C_\mathrm{DES}$ provided by the neural network are averaged over the face neighbour cells, weighted by the surface areas of the common faces.
Such approaches have been employed in the practical implementation of the dynamic Smagorinsky model and the adaptive $\ell^2-\omega$ DDES model \citep{yin2015dynamic} when determining a local dynamic model coefficient. 

In this work, the architecture of the neural network includes $2$ hidden layers with $10$ neurons per layer,
with the weight vector $\boldsymbol{w}$ having $203$ components.
The sensitivity of the architecture has been investigated in the previous study~\citep{zhang_ensemble-based_2022}, showing that the chosen network is adequate for providing converged model outputs.
The rectified linear unit (ReLU) activation function is used for the hidden layers, and the linear activation function is used for the output layer. 
The open-source library TensorFlow \citep{abadi2015tensorflow} is employed to construct the neural network.

\begin{figure}
    \centering
    \includegraphics[width=0.99\linewidth]{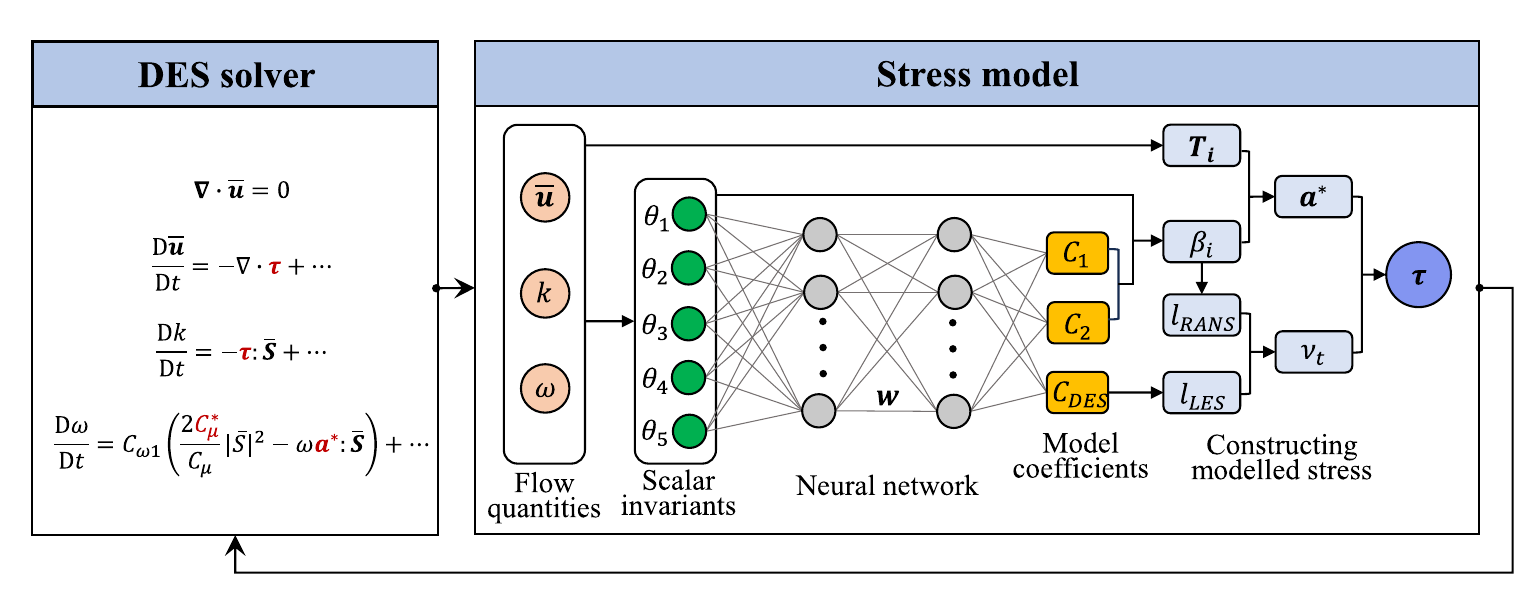}
    \caption{Schematic of the present data-driven EAS-DDES method with the model coefficients represented by neural networks.}
    \label{fig:NNsketch}
\end{figure}

The schematic of the present data-driven EAS-DDES method is illustrated in Figure \ref{fig:NNsketch}. 
Specifically, the governing equations of the flow quantities $\boldsymbol{\bar{u}}, k$, and $\omega$ are solved in the DES solver.
Then the tensor invariants $\theta_i$ and the tensor basis $\boldsymbol{T}_i$ are calculated. 
The invariants $\theta_i$ are fed into the neural network to obtain the model coefficients $C_1, C_2$, and $C_\mathrm{DES}$ as outputs. 
With $C_1, C_2$ and $\theta_j$ determined, the tensor basis coefficients $\beta_i$ can be obtained. 
Further, the coefficients $\beta_i$ and $C_\mathrm{DES}$ determine eddy viscosity $\nu_t$ through equations \eqref{DESnut}-\eqref{fd}. 
Meanwhile, $\beta_i$ and $\boldsymbol{T}_i$ determines the non-linear extra anisotropy $\bm{a^*}$ as equations \eqref{afull} and \eqref{tensorbasis}. 
Finally, the modelled stress $\boldsymbol{\tau}$ is constructed based on equation \eqref{tauij}, which is incorporated with the DES solver for solving the governing equations.
The present study employs the open-source finite-volume CFD toolbox -- OpenFOAM \citep{jasak2007openfoam} as the simulation platform.
The specific mapping from the input features ($\theta_i$) to the output coefficients ($C_1$, $C_2$, and $C_\mathrm{DES}$) for the neural network is determined by the parameter vector $\boldsymbol{w}$ (which consists of the weights and biases) of the neural network. 
The optimal $\boldsymbol{w}$ is obtained by ensemble Kalman-based training.

\subsection{Ensemble Kalman method for training neural network-based model\label{sec:training}}

The objective of the model training in the present work amounts to finding the optimal neural network weights $\boldsymbol{w}$ that 
minimize the model prediction error. 
The iterative ensemble Kalman method with adaptive stepping \citep{zhang_ensemble-based_2022} is employed for training the weights $\boldsymbol{w}$, which is briefly introduced here.

In the ensemble-based approach, the corresponding cost function reads \citep{zhang2020regularized}
\begin{equation}
\label{cost}
    J= \left\|\boldsymbol{y}-\mathcal{H}\left(\boldsymbol{w}_j^l\right)\right\|_{\gamma \boldsymbol{R}}^2 + \left\|\boldsymbol{w}_j^{l+1}-\boldsymbol{w}_j^l\right\|_{\boldsymbol{P}}^2 ,
\end{equation}
where $l$ is the iteration index, $j$ is the sample index,  $\|\cdot\|_{\boldsymbol{A}}$ indicates weighted norm (defined as $\|\boldsymbol{v}\|_{\boldsymbol{A}}^2=\boldsymbol{v}^{\mathrm{T}} \boldsymbol{A}^{-1} \boldsymbol{v}$ for a vector $\boldsymbol{v}$ with weight matrix $\boldsymbol{A}$), $\boldsymbol{P}$ is the model error covariance matrix indicating the uncertainties of the initial parameters, $\boldsymbol{R}$ is the data error covariance matrix, $\gamma$ is a scaling parameter for an adaptive adjustment of the step size at each iteration, and $\boldsymbol{y}$ is the training data subjected to the Gaussian noise $\epsilon \sim \mathcal{N}(0, R)$. 
For convenience, the operator $\mathcal{H}$ denotes a composition of the DES solver and the associated post-processing, which maps the weight parameters $\boldsymbol{w}$ to the observed quantities (e.g., mean velocity). 
In equation \eqref{cost}, the first term represents the
discrepancy between the model prediction and the training data. 
The second term is the regularization term, which penalizes large variation of $\boldsymbol{w}$ from the last iterative step.

By minimizing the cost function, the update scheme of the iterative ensemble Kalman method can be derived as
\begin{equation}
\qquad \boldsymbol{w}_j^{l+1}=\boldsymbol{w}_j^l+\boldsymbol{K}^l\left[\boldsymbol{y}-\mathcal{H}\left(\boldsymbol{w}_j^l\right)\right] \text{,}
\end{equation}
with $\boldsymbol{K}^l$ being the Kalman gain matrix calculated as
\begin{equation}
\boldsymbol{K}^l=\boldsymbol{S}_w^l (\boldsymbol{S}_y^l)^{\mathrm{T}}\left[\boldsymbol{S}_y (\boldsymbol{S}_y)^{\mathrm{T}}+\gamma^l \boldsymbol{R}\right]^{-1} \text{.}
\end{equation}
The square-root matrices $\boldsymbol{S}_w^l$ and $\boldsymbol{S}_y^l$ at each iteration are defined as
\begin{equation}
\begin{gathered}
\boldsymbol{S}_w^l=\frac{1}{\sqrt{N_j-1}}\left[\boldsymbol{w}_1^l-\hat{\boldsymbol{w}}^l, \boldsymbol{w}_2^l-\hat{\boldsymbol{w}}^l, \ldots, \boldsymbol{w}_{N_j}^l-\hat{\boldsymbol{w}}^l\right], \\
\boldsymbol{S}_y^l=\frac{1}{\sqrt{N_j-1}}\left[\mathcal{H}_1^l-\hat{\mathcal{H}}^l, \mathcal{H}_2^l-\hat{\mathcal{H}}^l, \ldots, \mathcal{H}_{N_j}^l-\hat{\mathcal{H}}^l\right] ,
\end{gathered}
\end{equation}
where $N_j$ is the sample size, and $\hat{\boldsymbol{v}}$ denotes the averaging of a vector $\boldsymbol{v}$ in the sample space. 
The parameter $\gamma^l$ is adjusted in an inner loop \citep{zhang_ensemble-based_2022}. 

As emphasized in \S \ref{sec:intro}, it is crucial to ensure a reasonable RANS/LES switching behaviour for a data-driven DES model.  
For the present DD-EAS-DDES approach, the switching behaviour is mainly controlled by $C_\mathrm{DES}$ and $C_\mu^*$, where $C_\mu^*$ is a function of tensor coefficients $\beta_i$ and the invariants $\theta_j$ as equation \eqref{Cmustar}. Here $\beta_i$ are explicitly solved from the algebraic equation \eqref{ARSM1} of modelled stress anisotropy with the model coefficients $C_1$ and $C_2$ determined. 
Given that $C_\mu^*$ directly controls the RANS length scale and $C_\mathrm{DES}$ determines the LES length scale, their relative magnitude imposes direct influence on the RANS-to-LES switching. 
Thus, the trained $C_1$, $C_2$, and $C_\mathrm{DES}$ are not allowed to deviate far away from baseline values, thereby maintaining a similar switching behaviour as the baseline EAS-DDES model. 
Hence, besides the flow quantities $q$ (e.g., mean velocity), the coefficients $C_1$, $C_2$, and $C_\mathrm{DES}$ normalized by the baseline values are also treated as the model outputs, with the ones viewed as the reference data, i.e.,
\begin{equation}
\begin{gathered} 
    \boldsymbol{y} =\left[ \alpha \frac{\langle {q} \rangle_\mathrm{data}}{q_\mathrm{ref}} , \ 1,\ 1,\ 1 \right], \\
    \mathcal{H}\left(\boldsymbol{w}_j^l\right) =\left[ \alpha \frac{\langle {q} \rangle_\mathrm{DES}}{q_\mathrm{ref}},\  \frac{\langle C_1 \rangle}{1.8},\ \frac{\langle C_2 \rangle}{5/9},\ \frac{\langle C_\mathrm{DES} \rangle}{0.12} \right], 
\end{gathered}
\end{equation}
where $q_\mathrm{ref}$ is the reference value of $q$ (\textit{e.g.,} $q_\mathrm{ref}$ is the bulk or freestream velocities when $q$ represents the mean velocity), and $\alpha$ is the weight between the flow quantities and the model coefficients. 
Here a large value of $\alpha = 100$ is set to ensure the dominant role of $\langle u \rangle$ in the training process. 
It has been verified that using values of $\alpha = 50$ or $200$ leads to negligible differences in the training results. 
This augmented observation is equivalent to adding a regularization term based on $C_1$, $C_2$, and $C_\mathrm{DES}$ in the cost function~\citep{zhang2021assimilation}.

Another benefit of treating the model coefficients as the training data is the ease of bounding their values to remain within physically meaningful ranges. 
Specifically, $C_\mathrm{DES}$ is supposed to be non-negative for the physical subgrid viscosity, and $C_1$ is required to be greater than unity for an ``return to isotropy'' behaviour of $\boldsymbol{a}$ \citep{pope2000turbulent}. 
Without the constraints on the model coefficients, these criteria can be violated during the training process.

The training algorithm involves the following steps.
\begin{enumerate}[label=(\roman*)]
    \item Pre-training. 
    The neural network weight vector $\boldsymbol{w}$ is pre-trained as $\boldsymbol{w^0}$ to preform as the baseline EAS-DDES model, i.e., $C_1 = 1.8$, $C_2= 5/9$, $C_\mathrm{DES} = 0.12$. 
    This can facilitate the training convergence and avoid non-physical coefficients when using conventional weight initialization methods.
    \item Sampling.
    Samples of the weight $\boldsymbol{w}$ are drawn from the normal distribution $\mathcal{N}(\boldsymbol{w^0}, \sigma^2)$. 
    Each realization $\boldsymbol{w}_j$ represents a specific neural-network-based turbulence model.
    \item Propagation.
    Each sample of $\boldsymbol{w}$ provides a specific mapping from the scalar invariants $\theta_i$ to the model coefficients $C_1, C_2$ and $C_\mathrm{DES}$. 
    As such, the turbulent stress is closed, and the DES solver is executed to acquire the statistics of the flow field, as described in \S \ref{sec:nnmodel}. 
    The model predictions in the observation space $\mathcal{H}_j^l$ are further obtained via post-processing (e.g., extracting mean velocities at specific points).
    \item Kalman update. With the predictions $\mathcal{H}_j^l$ and the corresponding training data $\boldsymbol{y}$, the weight vector $\boldsymbol{w}_j^l$ can be updated using the ensemble Kalman method described above.
\end{enumerate}
Steps (iii) and (iv) are repeated until one of the samples $\boldsymbol{w}_j^l$ reaches the convergence criteria. 
This sample of neural network weights is employed as the trained model in this work.
For details of the ensemble Kalman method and its practical implementation in turbulence modelling, please refer to \citet{zhang_ensemble-based_2022}. 
In the present work, the DAFI code \citep{strofer2021dafi} is used to implement the aforementioned training algorithm.

\section{Secondary flow in a square duct}
\label{sec:duct}
The proposed data-driven DES method is first tested for secondary flows in a square duct.
Explicit algebraic stress-based models offer a key advantage over linear eddy viscosity models by capturing Reynolds stress anisotropies, which are essential for representing turbulent secondary motions.
The square duct flow is one of the canonical secondary flows for testing the predictive performance of turbulence models.
The cross-stream secondary flow in a square duct is driven by the normal stress anisotropy $\tau_{yy} \neq \tau_{zz}$ and the shear stress $\tau_{yz}$
\citep{pettersson2002prediction, durbin2011statistical}.
Hence, we assess the ability of the proposed method in learning algebraic stress-based DES models for the secondary flow in a square duct. 
The baseline EAS-DDES model~\citep{liu2024explicit} has been used to simulate a square duct flow with bulk Reynolds number $Re_b = 2\delta u_b/\nu = 40000$, where $\delta$ is the duct half side, and $u_b$ is the bulk velocity. 
The prediction is compared with the DNS result \citep{pirozzoli2018turbulence}, showing substantially better accuracy than traditional LEV-DDES models, e.g., the $\ell^2-\omega$ DDES \citep{reddy2014ddes} and the IDDES-SST \citep{gritskevich2012development} models. 
Here we show that the proposed method can further improve the predictive accuracy by providing the neural network-based model coefficient closures.

\subsection{Case set-up and numerical methods}
\label{sec:casesetup_square}

The square duct flow with $Re_b = 17800$ is chosen to train the model based on the DNS data of \citet{pirozzoli2018turbulence}. 
The computational domain length is $ L_x = 20\delta$ in the streamwise direction. 
The structural mesh consists of 1.024 million cells, with 160 cells in the streamwise direction and 80 cells in the duct side direction. 
The grid is uniformly distributed in the streamwise direction, with the mean wall-unit normalized size $\upDelta x^+ \approx 30$. 
The grid spacing is progressively refined in the wall-normal direction, with the cell size of the near-wall first layer being $\upDelta y^+_1 < 1$. 
The maximum cell spacing in cross-stream directions is $\upDelta y^+_{\max} = \upDelta z^+_{\max} \approx 15$. 
The streamwise boundaries are set as periodic, and the other boundaries are no-slip walls. 
A global pressure gradient drives the flow to maintain a desired value of $Re_b$. 

The flow statistics are obtained by averaging over time, the homogeneous streamwise direction, and the four quadrants. 
Due to the weakness of the secondary flow, the required averaging time interval $\upDelta t_{av}$ to achieve statistical convergence is extremely long. 
To account for the influence of streamwise extent of the domain in the statistical convergence \citep{pirozzoli2018turbulence}, the effective time averaging intervals are set as $\upDelta t_{av}^* =  \upDelta t_{av} L_x/(6\delta) \approx 200 \delta/u_\tau $ to achieve the statistical convergence of the first-order moment of the velocity field during the training process.

The numerical scheme achieves second-order accuracy in both time and space. 
The central linear scheme is utilized for momentum convection.
For the convection of turbulence variables, i.e., $k$ and $\omega$, the central linear scheme with the Sweby limiter is applied to ensure boundedness. 
Diffusion terms are treated using central linear interpolation for face values and Gauss’s theorem for surface integrals. 
Time discretization is performed using the second-order implicit Euler method. 
The time-step size is dynamically adjusted to maintain the maximum Courant–Friedrichs–Lewy (CFL) number below 0.5. 
The coupling between pressure and velocity is managed through the PIMPLE algorithm.
A preconditioned conjugate gradient solver is employed for the pressure equation, while other transport equations are solved using a preconditioned biconjugate gradient method.

The training data set consists of the mean velocities $\langle \bar{u}_x \rangle$, $\langle \bar{u}_y \rangle$ and $\langle \bar{u}_z \rangle$ in the whole computational domain, with $\langle \bar{u}_y \rangle$ and $\langle \bar{u}_z \rangle$ being scaled up by a factor of 50 to have a similar maximum magnitude with $\langle \bar{u}_x \rangle$. 
The training data is weighted by the local cell volume. 
During model training with the ensemble Kalman method, 16 samples are drawn in this case, and convergence of the cost function is achieved within approximately 5 iterative steps. 
In each iteration step, the computational cost for each sample is approximately 1000 CPU hours (equivalent to the computational cost of the baseline or trained DD-EAS-DDES models in this case), resulting in a total training cost of 80,000 CPU hours.

\subsection{Prediction of velocity and modelled stress}

\begin{figure}
    \centering
    \includegraphics[width=0.99\linewidth]{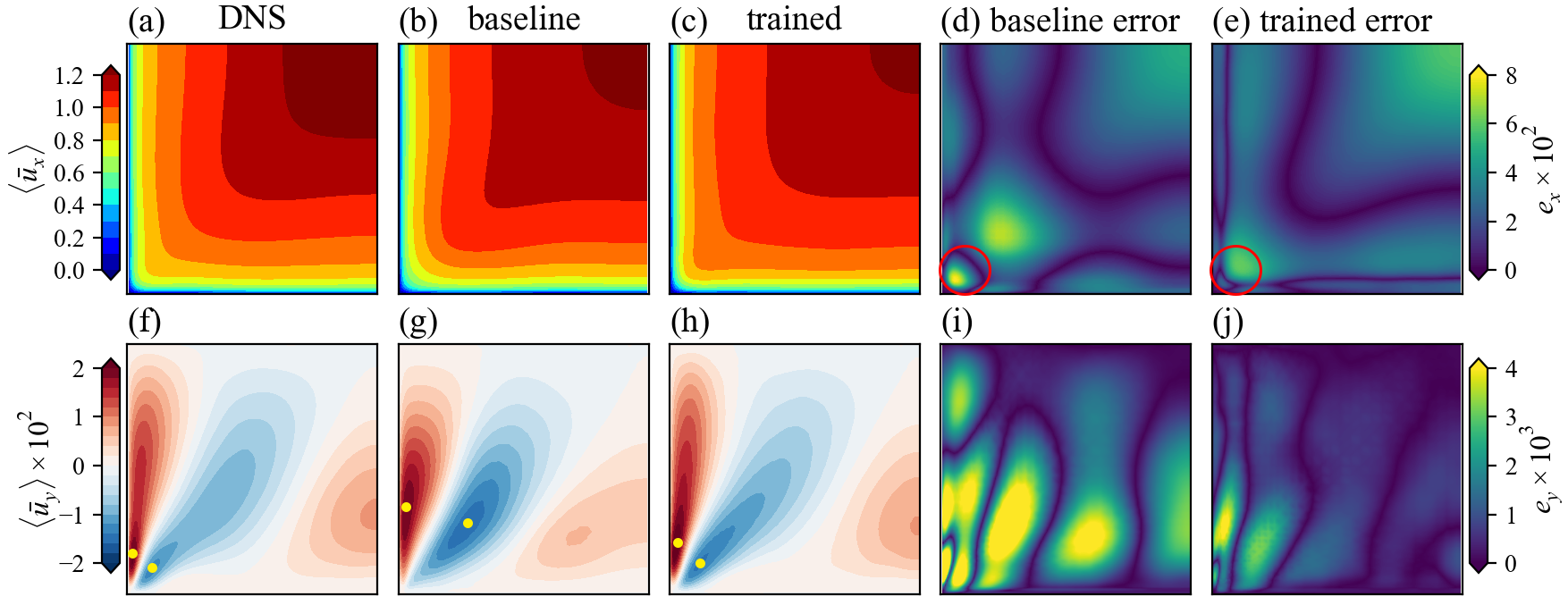}
    \caption{Mean velocity in (a-c) $x$- and (f-h) $y$-directions and absolute prediction error contours in (d, e) $x$- and (i, j) $y$-directions in the third quadrant ($y<0$, $z<0$) for the square duct.}
    \label{fig:square_duct_U_contour}
\end{figure}

The mean velocity contours predicted by the baseline and trained EAS-DDES models are compared with the DNS results in figures \ref{fig:square_duct_U_contour} (a-c, f-h). 
Only the third quadrant is shown due to the flow 
symmetry. 
The accuracy of the trained model outperforms the baseline model in both the main flow ($\langle \bar{u}_x \rangle$) and the secondary flow ($\langle \bar{u}_y \rangle$).
Figure \ref{fig:square_duct_U_contour} (d, e, i, j) shows the absolute error of the predictions by both models with respect to the DNS data, defined as 
 \begin{equation}
 \label{error}
     e_{x,y} = | \langle \bar{u}_{x,y}^{\text{trained}} \rangle - \langle \bar{u}_{x,y}^{\text{data}} \rangle | \text{.}
 \end{equation}
The absolute error of the trained model is significantly reduced compared to the baseline model. 
For the streamwise mean velocity, the improvement of the trained model is significant near the corner of the duct, as marked by the red circles in figure \ref{fig:square_duct_U_contour}(d, e). For the prediction of the secondary flow, the trained model performs remarkably better in the entire domain. 
It is observed that the peaks of cross-stream velocity (including both positive and negative values) predicted by the trained model are located near the corner regions in closer agreement with the DNS data compared to the baseline model, as marked by the yellow symbols in figure \ref{fig:square_duct_U_contour}(f-h).

\begin{figure}
    \centering
    \includegraphics[width=0.99\linewidth]{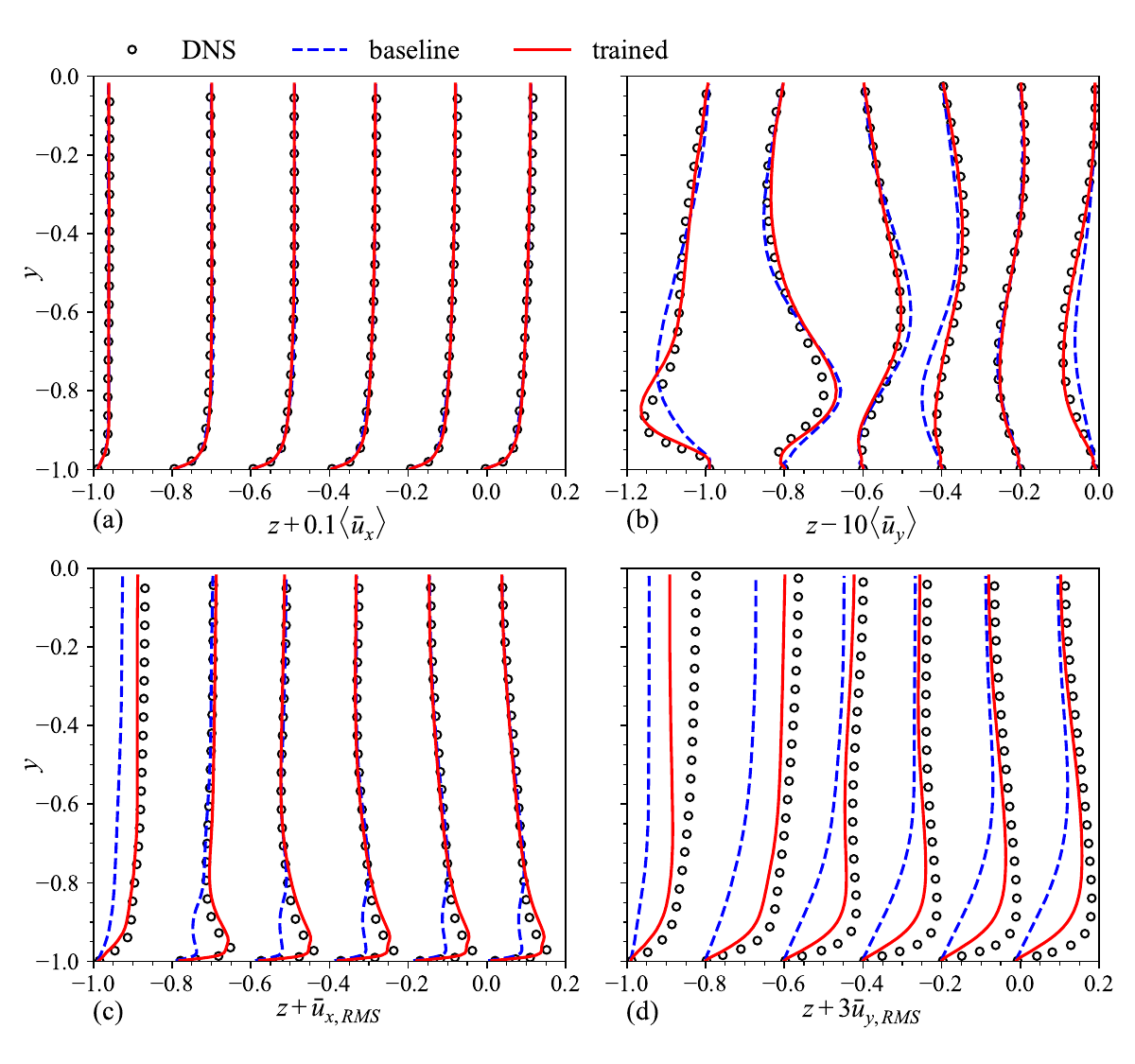}
    \caption{Comparison of the mean and RMS velocity profiles between the DNS, the baseline EAS-DDES model, and the trained DD-EAS-DDES model at $z = -0.99$, $-0.8$, $-0.6$, $-0.4$, $-0.2$, $0$ for flows in the square duct.}
    \label{fig:square_duct_U_lines}
\end{figure}

Figure \ref{fig:square_duct_U_lines}(a,b) presents the mean velocity profiles at different cross-stream locations. 
It can be observed that there exist negligible differences in the streamwise mean velocity between the DNS results and the predictions by different models. 
For the secondary flow predictions, a substantial improvement in prediction accuracy is observed for the trained model compared to the baseline model, especially in the near-wall region, e.g., $z = 0.99$. 
This indicates that the trained DES model is effective in both the RANS and the LES branches in this case.

As for the root-mean-square (RMS) velocity profiles illustrated in figure \ref{fig:square_duct_U_lines}(c,d), the trained model provides significantly better predictions than the baseline model in both the streamwise and the cross-stream directions. Overall, the baseline model under-predicts the RMS velocities in the near-wall regions, while the trained model largely compensates for this due to the enlarged LES region. This will be further demonstrated in \S \ref{sec:switching}.

\begin{figure}
    \centering
    \includegraphics[width=0.99\linewidth]{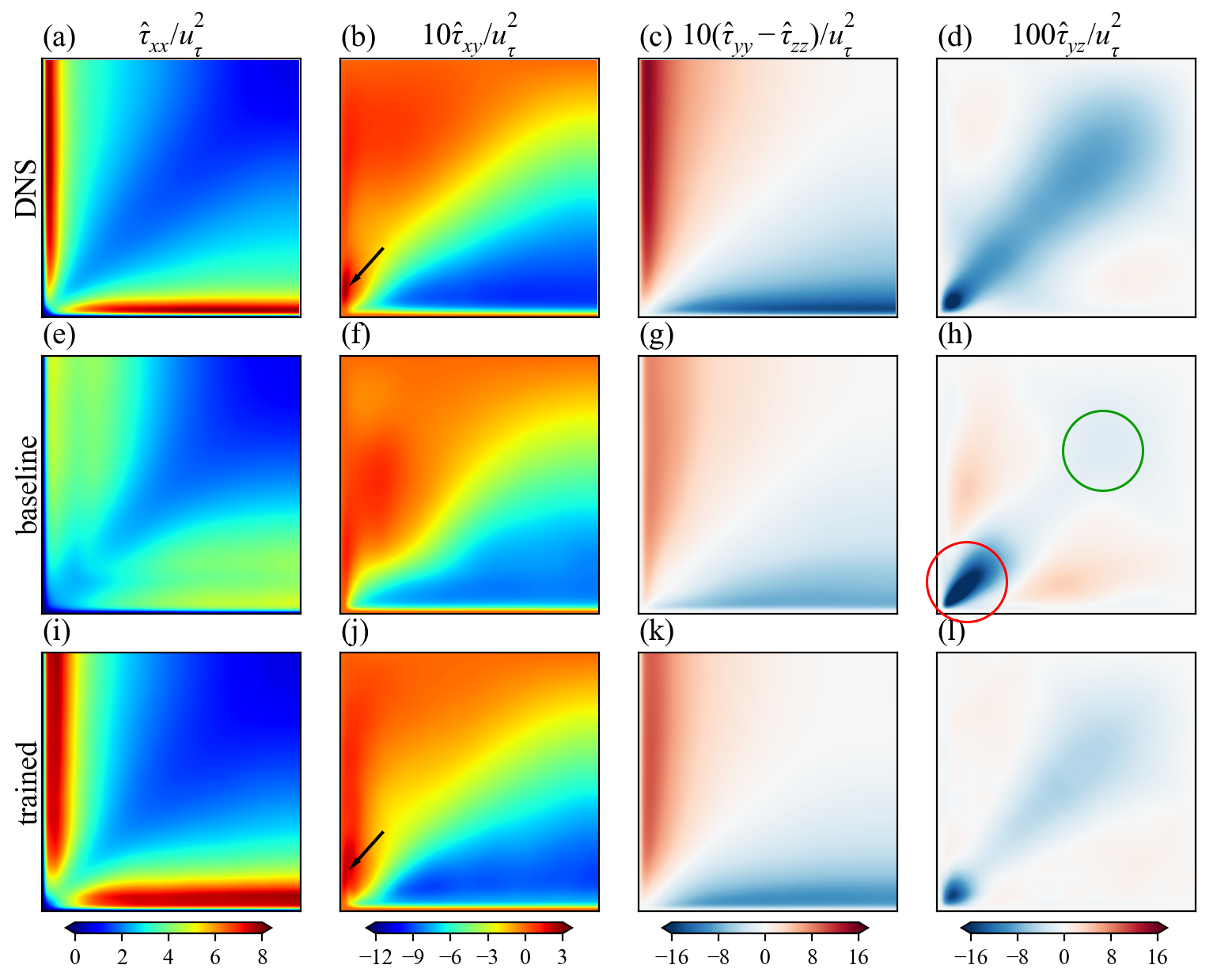}
    \caption{Comparisons of the normalized total stress between the (a-d) DNS, (e-h) the baseline EAS-DDES model, and (i-l) the trained DD-EAS-DDES model in the third quadrant ($y<0$, $z<0$) of the square duct.}
    \label{fig:square_duct_tau}
\end{figure}

To clarify the underlying mechanism of the improved velocity predictions, the normalized total stress $\hat{\boldsymbol{\tau}} / u_\tau^2$ is illustrated in figure~\ref{fig:square_duct_tau}, where $u_\tau$ is the wall-friction velocity from the DNS. 
For the DES, $\hat{\boldsymbol{\tau}}$ is defined as the summation of the resolved and the modelled parts as
\begin{equation}
    \label{tau_tot}
    \hat{\boldsymbol{\tau}} \equiv \langle \boldsymbol{u}'\boldsymbol{u}' \rangle   =  \langle \bar{\boldsymbol{u}}' \bar{\boldsymbol{u}}' \rangle + \langle \boldsymbol{\tau} \rangle \text{.}
\end{equation}
For the streamwise normal stress $\hat{\tau}_{xx}$, the baseline model underestimates the peak value and overestimates the thickness of the high $\hat{\tau}_{xx}$ region, while the prediction of the trained model is in better agreement with the DNS results.
Similarly, for the streamwise shear stress $\hat{\tau}_{xy}$, the trained model provides better agreement with the DNS than the baseline model. 
Notably, the trained model reproduces the small positive peak near the corner observed in the DNS, as marked by the black arrows in figure \ref{fig:square_duct_tau}(b, j), while the baseline model fails to capture it.

The inherent nature of the secondary flow can be illustrated by the mean streamwise vorticity equation
\begin{equation}
\label{omega_x}
       \left( \langle {u}_y \rangle \frac{\partial  }{\partial y}+\langle {u}_z \rangle \frac{\partial }{\partial z} \right) \langle {\omega}_x \rangle 
    - \nu \left(\frac{\partial^2 }{\partial y^2}+\frac{\partial^2 }{\partial z^2}\right) \langle {\omega}_x \rangle 
=
 \left(\frac{\partial^2 }{\partial z^2}-\frac{\partial^2 }{\partial y^2}\right) \hat{\tau}_{yz}
+\frac{\partial^2}{\partial y \partial z}\left(\hat{\tau}_{yy} - \hat{\tau}_{zz} \right).
\end{equation}
The four terms in equation \eqref{omega_x} represent the effects of convection, diffusion, secondary turbulent shear stress, and the turbulent normal stress anisotropy, respectively.
The normal stress anisotropy $\left(\hat{\tau}_{yy} - \hat{\tau}_{zz} \right)$ has been widely recognized as driving the streamwise vorticity due to its dominant magnitude. 
On the other hand, it is evident from numerical simulations \citep{pirozzoli2018turbulence} that the secondary turbulent shear stress term in the mean streamwise vorticity equation has a similar magnitude to the normal stress anisotropy term. 
Moreover, it is found that non-zero turbulence production is mainly associated with significant secondary shear stress. 
Given this, we plot both the normal stress anisotropy $(\hat{\tau}_{yy} - \hat{\tau}_{zz})$ and the secondary turbulent shear stress $\hat{\tau}_{yz}$, as shown in the third and fourth columns of figure \ref{fig:square_duct_tau}. 
The difference between the baseline and trained model predictions in $(\hat{\tau}_{yy} - \hat{\tau}_{zz})$ is marginal, with the predictions by the trained model in slightly better agreement with the DNS result. 
As for $\hat{\tau}_{yz}$, it is obvious that the trained model outperforms the baseline model. 
Specifically, the distribution of $\hat{\tau}_{yz}$ for the trained model prediction is similar to the DNS results.
However, the baseline model significantly overestimates the magnitude of $\hat{\tau}_{yz}$ in the corner region (red circle), and underestimates it in the central region of the duct (green circle). 
These improvements in modelled stress lead to better velocity predictions of the secondary flow with the trained model than the baseline model.

\subsection{Switching behaviours and model coefficients\label{sec:switching}}

\begin{figure}
    \centering
    \includegraphics[width=0.99\linewidth]{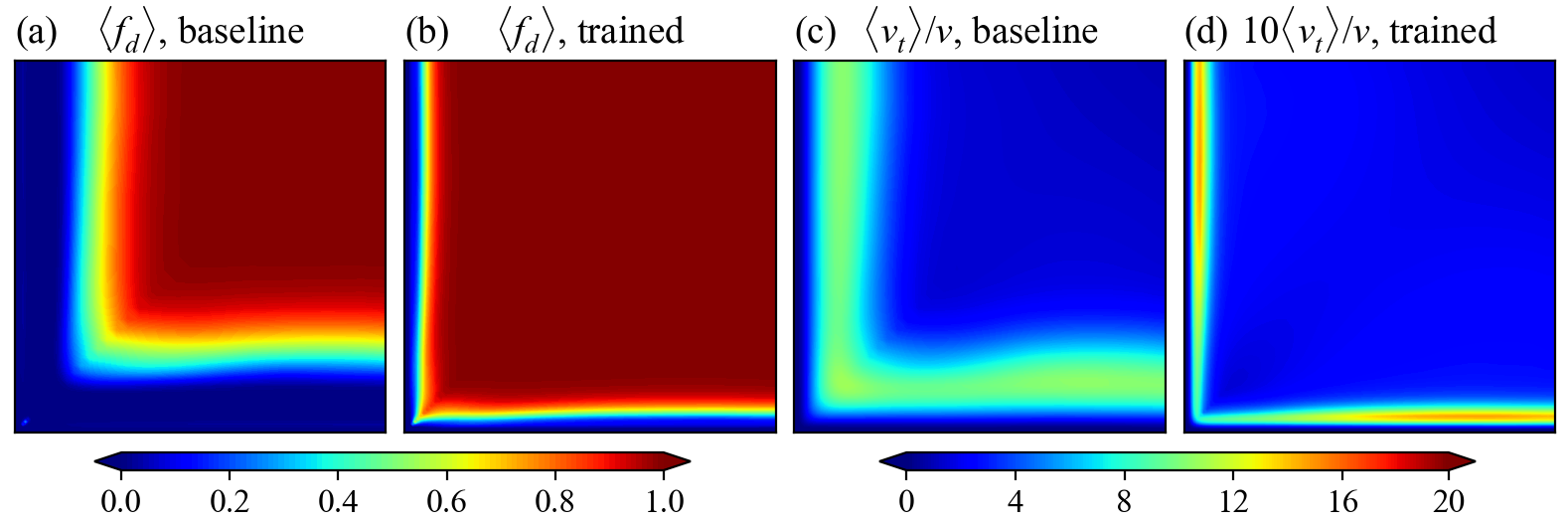}
    \caption{Comparisons of the shielding function $f_d$ and the turbulent viscosity ratio $\nu_t/\nu$ between the baseline EAS-DDES model and the trained DD-EAS-DDES model.}
    \label{fig:square_duct_fd_contour}
\end{figure}

Figure \ref{fig:square_duct_fd_contour} compares the switching behaviours between the RANS and LES branches for the baseline and the trained models. 
Specifically, the mean shielding function $\langle f_d \rangle$ is shown in figures \ref{fig:square_duct_fd_contour}(a, b), and the mean eddy viscosity ratio $\langle \nu_t \rangle / \nu$ is shown in figures \ref{fig:square_duct_fd_contour}(c, d). 
Strictly speaking, the shielding function $\langle f_d \rangle$ in DDES ensures a shielded near-wall RANS region to prevent the GIS issue. 
Outside the shielded region, the RANS and LES length scales ( $\ell_{RANS}$ and $\ell_{LES}$) are compared locally to determine the RANS/LES mode, as described in \S \ref{sec:modelDDES}. 
Usually, $f_d = 1$ alone does not sufficiently indicate a LES region. 
However, in the present square duct flow simulations, the mesh is sufficiently fine outside the shielded region ($\upDelta x^+ \approx 30$ and $\upDelta y^+_{\max} = \upDelta z^+_{\max} \approx 15$), ensuring $\ell_{LES} < \ell_{RANS}$ in these regions. 
Therefore, $f_d$ is sufficient to indicate the switching behaviour in this case.

It can be seen that the trained model performs a much larger region of the LES branch as indicated with $\langle f_d \rangle = 1$. 
The near-wall RANS region is much thinner than that of the baseline model. 
This leads to a relatively small eddy viscosity ratio for the trained model, compared to the baseline.
Note that $\langle \nu_t \rangle / \nu$ for the trained model in figure \ref{fig:square_duct_fd_contour}(d) is scaled by a factor of 10 for clarity. 
These facts indicate that the trained model enhances performance by enlarging the LES region to resolve more turbulent structures.

\begin{figure}
    \centering
    \includegraphics[width=0.99\linewidth]{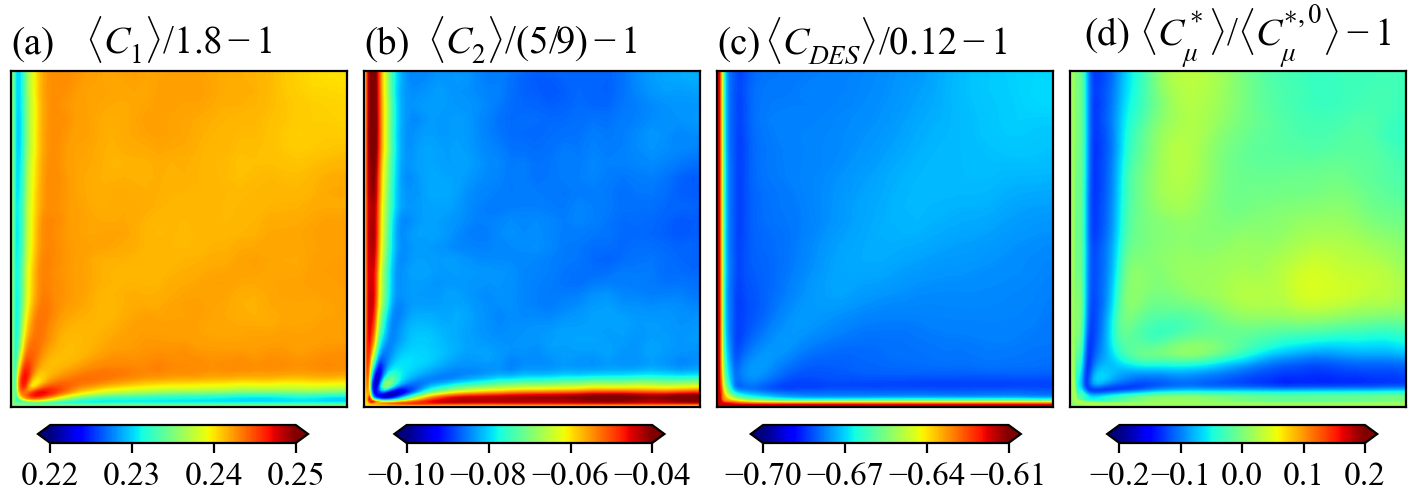}
    \caption{Mean values of the model coefficients $\langle C_1 \rangle$, $\langle C_2 \rangle$, $\langle C_\mathrm{DES} \rangle$ and $\langle  C_\mu^* \rangle$ of the trained DD-EAS-DDES model in the third quadrant ($y<0$, $z<0$) of the square duct. The coefficients are normalized by the baseline values.}
    \label{fig:square_duct_c1c2cdes_contour}
\end{figure}

To explore the underlying reasons for the trained model's behaviour, the model coefficients $C_1$, $C_2$, and $C_\mathrm{DES}$, and the calculated $C_\mu^*$ are examined. 
The relative variations of these model coefficients compared to the baseline values are presented in figure \ref{fig:square_duct_c1c2cdes_contour}. 
As shown in figure \ref{fig:square_duct_c1c2cdes_contour}(d), $C_\mu^*$ in the near-wall region is lower than the baseline value, which is determined by the variation of $C_1$ and $C_2$ from the baseline value. 
This coefficient directly controls $f_d$ as equation \eqref{fd}. 
That is, a smaller $C_\mu^*$ leads to a larger value of $f_d$, consequently, a thinner shielded RANS region. 
Additionally, $C_\mu^*$ and $C_\mathrm{DES}$ jointly control the production term of the $k$ transport equation as \eqref{toteq}. 
The simultaneous decrease of $C_\mu^*$ and $C_\mathrm{DES}$ reduces the modelled kinetic energy, which further increases $f_d$. 
Therefore, both the variations of $C_\mathrm{DES}$ and $C_\mu^*$ result in enlarging the LES region and resolving more turbulent structures.

\begin{figure}
    \centering
    \includegraphics[width=0.99\linewidth]{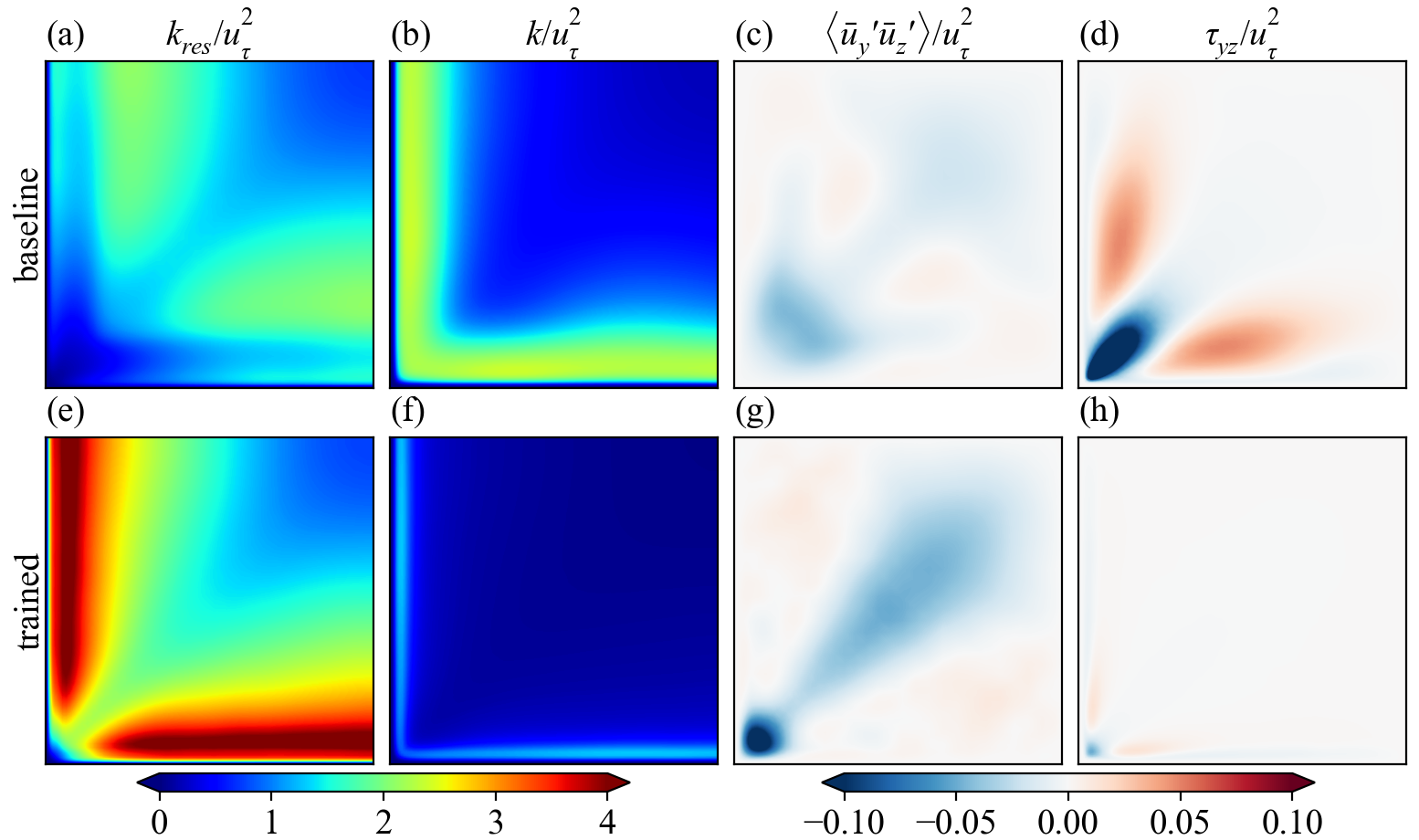}
    \caption{Comparisons of the mean resolved and modelled parts of TKE and secondary shear stress between the baseline EAS-DDES model and the trained DD-EAS-DDES model in the third quadrant ($y<0$, $z<0$) of the square duct.}
    \label{fig:square_duct_aij}
\end{figure}

To investigate the effects of the trained model coefficients, we present the resolved and modelled parts of the turbulent stress as shown in figure~\ref{fig:square_duct_aij}.
From figure~\ref{fig:square_duct_aij}(a, b, e, f), the increasing trend of resolved TKE portions for the trained model can be seen.
For the baseline model, the modelled TKE is primarily distributed in the near-wall region, and the resolved part is distributed away from the wall, with both exhibiting a similar maximum magnitude. 
While for the trained model, TKE is almost fully resolved, with the modelled part being small. 

As discussed above, the secondary flow is mainly driven by the turbulent normal stress anisotropy $(\hat{\tau}_{yy} - \hat{\tau}_{zz})$ and the secondary turbulent shear stress $\hat{\tau}_{yz}$. 
Specifically, the predictions of the baseline and the trained model have major differences in $\hat{\tau}_{yz}$ (see the last column of figure \ref{fig:square_duct_tau}). 
Figures \ref{fig:square_duct_aij}(c, d, g, h) illustrate the resolved and modelled parts of $\hat{\tau}_{yz}$ for the baseline and trained models. 
For the baseline model, the modelled part accounts for a large proportion of $\hat{\tau}_{yz}$, while the resolved part is relatively small. 
The spatial distribution of $\tau_{yz}$ is very similar to the total one $\hat{\tau}_{yz}$ (figure \ref{fig:square_duct_tau}(h)), and the deviation from the DNS result mainly comes from the modelling error. 
In contrast, the major contribution of $\hat{\tau}_{yz}$ is from the resolved part for the trained model, which is similar to the distribution of the DNS result (figure \ref{fig:square_duct_tau}(d)). 
This further confirms that the better performance of the trained model is mainly attributed to the increase in the resolved turbulence.  

\begin{figure}
    \centering
    \includegraphics[width=0.75\linewidth]{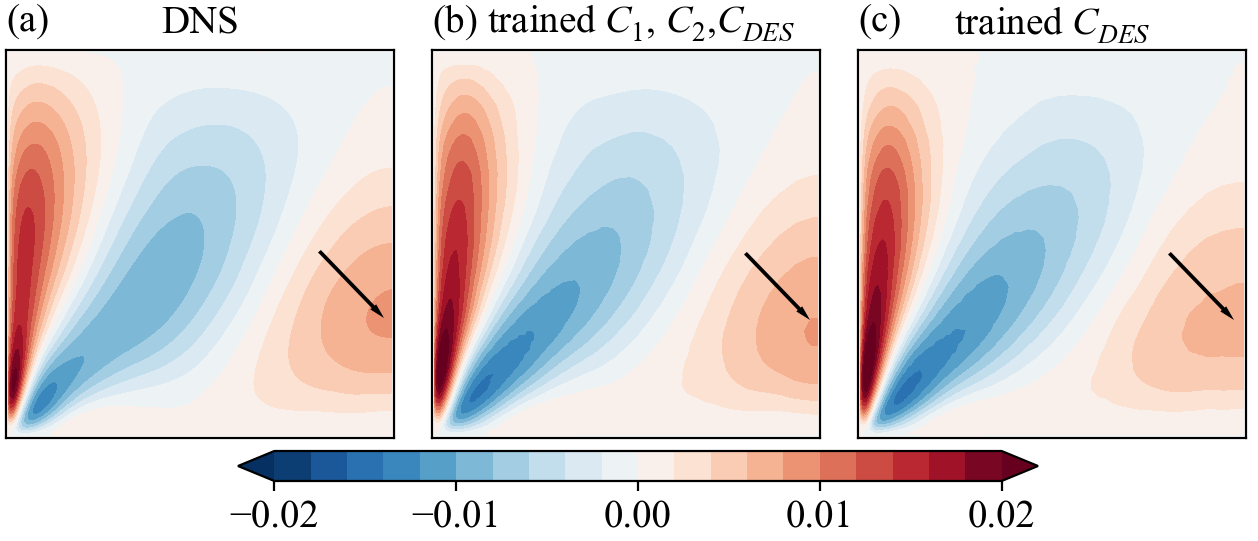}
    \caption{Comparison of the mean secondary velocity $\langle \bar{u}_y \rangle$ between the DNS, the fully trained DD-EAS-DDES model and the DD-EAS-DDES model with trained $C_\mathrm{DES}$ in the third quadrant ($y<0$, $z<0$) of the square duct.}
    \label{fig:square_duct_Uy_contour_cdesTrained}
\end{figure}

The variations of the coefficients $C_1$ and $C_2$ have two main effects on the model prediction. 
The first is to alter the switching location between the RANS and LES regions by modifying $C_\mu^*$. 
The second one is to alter the non-linear extra anisotropy $\boldsymbol{a^*}$ by modifying the expressions of the tensor basis coefficients $\beta_i$ as listed in the Appendix \ref{sec:appA}. 
To assess the contribution of the second effect, a new simulation is conducted by using the baseline $C_1$, $C_2$ values, and the trained $C_\mathrm{DES}$ function. 
This can lead to a similar $f_d$ distribution as the fully trained model, 
while the functions of $\beta_i$ with respect to the scalar invariants $\theta_j$ are unchanged because of using the baseline $C_1$ and $C_2$ values. 
The results of the mean secondary velocity $\langle \bar{u}_y \rangle$ are shown in figure \ref{fig:square_duct_Uy_contour_cdesTrained}. 
Overall, the predictions from the partially trained case are similar to the fully trained case.
A difference can be observed in the positive $\langle \bar{u}_y \rangle$ region near the central line ($z=0$), as pointed by the black arrows, where the fully trained model has slightly better performance than the partially trained one. 
Hence, the modifications of $\beta_i$ functions with the trained $C_1$ and $C_2$ have positive effects on the model performance, although the impact is weaker than the modification of LES/RANS switching location via the trained $C_\mathrm{DES}$.

\subsection{Generalization in Reynolds number}

\begin{figure}
    \centering
    \includegraphics[width=0.99\linewidth]{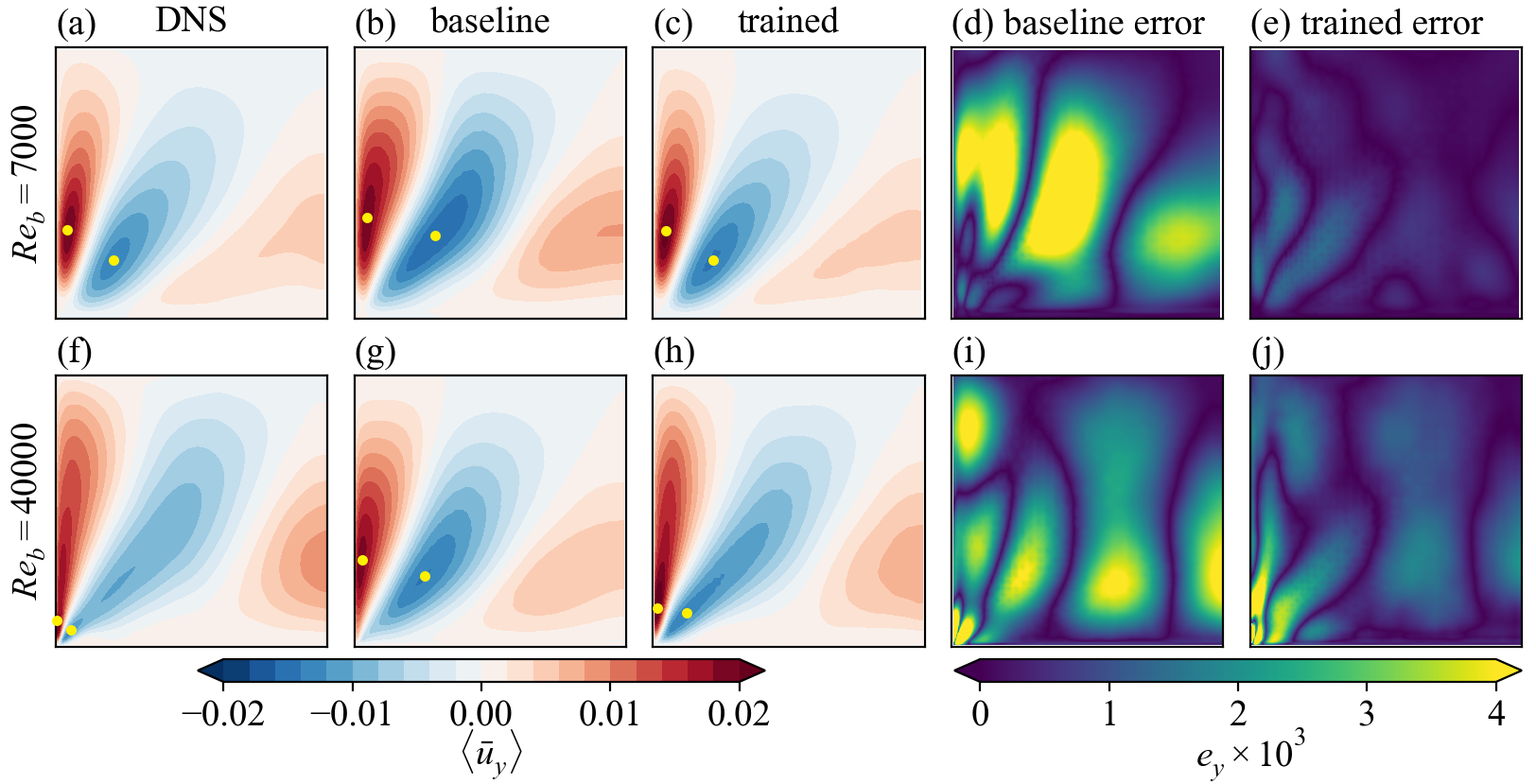}
    \caption{(a-c, f-h) The mean secondary velocity $\langle \bar{u}_y \rangle$ of the DNS result, the predictions by the baseline and trained EAS-DDES models, and (d, e, i, j) the absolute errors for the baseline and trained models in the third quadrant ($y<0$, $z<0$) of the square duct flow with (a-e) $Re_b = 7000$ and (f-j) $Re_b = 40000$. 
    The model is trained in the square duct flow with $Re = 17800$.}
    \label{fig:square_duct_Uy_contour_Re40000}
\end{figure}

The generalizability of the trained model is further assessed in two square duct flows of $Re_b = 7000$ and 40000.
The mesh is the same as the trained case, which means the grid resolution relative to turbulence scales varies due to the variation of the Reynolds number, with a larger proportion of turbulent scales being modelled at a higher Reynolds number.
As such, the model's generalizability can also be assessed in the grid resolution with respect to the turbulence scales.
The secondary flow predictions of the baseline and the trained models are compared to the DNS result in figure \ref{fig:square_duct_Uy_contour_Re40000}.
A significant improvement in the accuracy of the trained model is observed for both cases, which is similar to the case of $Re_b = 17800$ as shown in figure \ref{fig:square_duct_U_contour}(f-h). 
The predicted locations of the positive and negative peaks of $\langle \bar{u}_y \rangle$ (as marked by the yellow symbols) are closer to the duct corner with the increase of Reynolds number, and the positive $\langle \bar{u}_y \rangle$ near the central line ($z = 0$) is better predicted than the baseline model.
Compared with the baseline model, the trained model's absolute errors are remarkably reduced for both cases in the whole domain.
Conclusively, the trained model can be well generalized to flows with this different Reynolds number.

\begin{figure}
    \centering
    \includegraphics[width=0.99\linewidth]{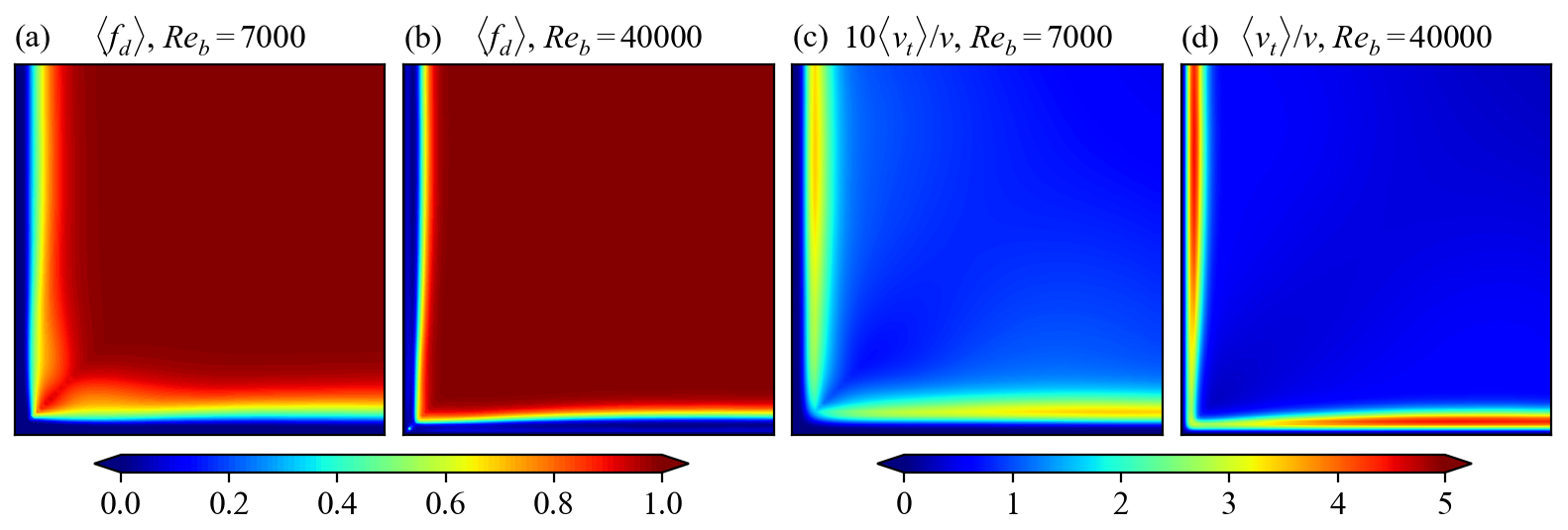}
    \caption{Comparisons of the time-averaged DDES shielding function $f_d$ and the turbulent viscosity ratio $\nu_t/\nu$ between the $Re_b = 7000$ and the $Re_b = 40000$ cases simulated by the trained DD-EAS-DDES model. 
    The model is trained in the square duct flow with $Re = 17800$. }
    \label{fig:square_duct_fd_contour_Re40000}
\end{figure}

The switching behaviours of the trained model can be well generalized to the cases with different Reynolds numbers.
It is supported in figure \ref{fig:square_duct_fd_contour_Re40000}, which compares the switching behaviours of the $Re_b = 7000$ and $Re_b = 40000$ cases. 
Both cases are simulated based on the DD-EAS-DDES model trained by the $Re_b = 17800$ case. 
The distribution of $f_d$ is almost unchanged for the two different Reynolds numbers, while the eddy viscosity ratio $\langle \nu_t \rangle/\nu$ in the near-wall RANS region becomes significantly larger for higher Reynolds numbers. 
For clarity, $\langle \nu_t \rangle/\nu$ is scaled by a factor of 10 for the $Re_b = 7000$ case in figure \ref{fig:square_duct_fd_contour_Re40000}(c).
This indicates a larger proportion of modelled stress at higher Reynolds numbers.

\section{Separated flows over a bump}
\label{sec:bump}

We further test the proposed data-driven detached eddy simulation method in the flow over a bump, which is another challenging case for either the RANS or DES models. 
\citet{deck2012recent} classifies the flow separation into three categories. 
The first one is the separation fixed by the geometry, such as the separated flow over a backward-facing step. 
The second and third categories are the separation induced by a pressure gradient on a curved surface. 
The second one has an inflow boundary layer thickness $\delta/H \ll 1$, where $H$ is the characteristic height of the curved surface, while the third one has $\delta/H = O(1)$. 
For the third category, the separation is strongly influenced by the dynamics of the incoming turbulent boundary layer. 
The turbulence statistics inside the boundary layer are strongly affected by the pressure gradient over the curved surface.
The investigated flow over the bump belongs to the third category.
The favourable pressure gradient (FPG) in the upstream side of the bump accelerates the mean flow while reducing the turbulent intensity, which may cause relaminarization \citep{balin2021direct}.
On the other hand, the adverse pressure gradient (APG) in the downstream side of the bump decelerates the mean flow, enhancing the turbulence and causing the flow separation. 
These features pose significant challenges for turbulence modelling of the flow over a bump, especially in capturing the separation point, the extent of the separated region, and the subsequent reattachment behaviour accurately \citep{durbin2018some}. 
The wall-resolved LES (WRLES) datasets by \citet{matai2019large} are used to evaluate the capability of the DD-EAS-DDES model in such flows. 
In the datasets, the LES results of flow over a parametric set of bumps with different crest heights are available, which are used for the generalization test of data-driven DES models.

\subsection{Case set-up and numerical methods}

\begin{figure}
    \centering
    \includegraphics[width=0.99\linewidth]{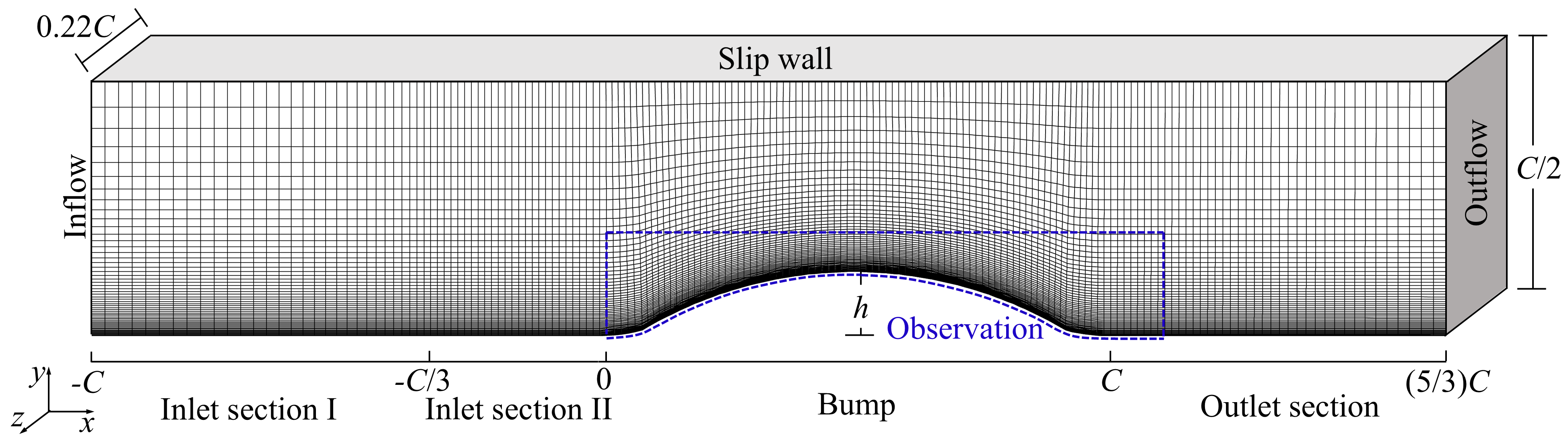}
    \caption{The computation domain and mesh on the $x$-$y$ plane for the separated flow over a bump.}
    \label{fig:bump_mesh}
\end{figure}

The flow over the bump with $h = 38$ mm is chosen to train the DES model. 
The computational domain and mesh on the $xy$ plane are shown in figure \ref{fig:bump_mesh}.
The length of the bump is $C = 305$ mm, which consists of a circular arc with a chord length of $(5/6)C$ and convex fillets with a radius of $0.323C$ added before and after the arc.
The Reynolds number of the bump case is $Re_c = 2 \times 10^5$ based on the free-stream velocity and the bump length.
Before the bump, the inlet section consists of two parts. 
The inlet section I with a length of $(2/3)C$ is added to ensure adequate development of synthetic turbulence introduced at the inlet. 
The inlet section II has a length of $C/3$, which is consistent with the LES configuration.
The outlet section length after the bump is $(2/3)C$. 
The height of the domain is $C/2$, and the width is $0.22C$. 
The computational domain is the same as the LES case \citep{matai2019large} except for the inclusion of the inlet section I.

The structural mesh consists of 0.42 million cells, with 175, 60, and 40 cells in the streamwise, wall-normal, and spanwise directions, respectively. 
In the streamwise direction, the grid is uniformly distributed for inlet section I and the bump region, with the cell size $\upDelta x/C = 0.013$ and $\upDelta x^+ \approx 100$. 
For the inlet section II and the outlet section, the grid size is stretched in the streamwise direction by a factor of 2 towards the inlet and the outlet, respectively. 
In the wall-normal direction, the grid spacing is progressively refined towards the bottom wall, with the cell size of the near-wall first layer satisfying $\upDelta y^+_1 < 1$. 
The grid is uniformly distributed in the spanwise direction, with the normalized size $\upDelta z/C = 0.0055$ and $\upDelta z^+ \approx 40$.

The periodic boundaries are set for the spanwise direction. 
The slip and no-slip boundaries are set for the top and bottom walls, respectively. 
The outlet is imposed with fixed pressure, zero velocity gradient, and non-reverse flow.
At the inlet, the velocity follows LES practice by adding synthetic turbulence to the mean velocity profile to obtain the instantaneous velocity.
The synthetic turbulence is constructed by the synthetic digital filter method (SDFM) \citep{klein2003digital}. 
It employs the digital filtering of random data to reproduce specified second-order (single-point) statistics and the autocorrelation functions of the velocity. 
The SDFM has been widely used in generating desired inflow turbulence in eddy-resolving simulations \citep{wu2017inflow}.
The required mean velocity, Reynolds stress, and integral length profiles are provided by a precursor RANS simulation of a zero-pressure-gradient flat plate. 
The streamwise location $x_0$ with the same boundary layer thickness as the LES inlet is prescribed, and the profiles at $x = x_0 - (2/3)C$ are extracted for generating the inlet instantaneous velocity. 
For the turbulence properties, the $\omega$ profile is also extracted from the precursor RANS simulation of the flat plate. 
In the DES, $k$ represents the modelled part of TKE, which is set to zero at the inlet, and synthetic turbulence is introduced to represent the resolved TKE.
The subgrid TKE is underestimated near the inlet, while it can be developed to a physical value after the inlet section I. 

The numerical schemes are the same as the square duct flow case described in \S \ref{sec:casesetup_square}. 
The time-step size is fixed at a value with a maximum CFL number of approximately 0.5. 
The flow statistics are obtained by averaging over time and spanwise direction, with the averaging time interval $\upDelta t_{av} \approx 8 C/u_b$.

In this case, the flow separation and reattachment are of primary interest.
Given this, the training data are selected between the beginning of the bump ($x/C = 0$) and the reattachment point  ($x/C = 1.1$) based on the LES in the $x$ direction. 
The range in the $y$ direction is $y < 0.2C$. 
The observation region is shown as the blue dashed lines in figure \ref{fig:bump_mesh}.
In this region, flow data are extracted along 23 parallel vertical lines with an interval of $0.05C$ in the $x$-direction. 
For all cells along these lines, the mean streamwise velocity divided by the local wall distance, $\langle \bar{u}_x \rangle / d_w$, is used as the training data. 
Here, $d_w$ is introduced to compensate for the small velocity values near the wall, as suggested by \citet{matai2019large}. 
The training data is also weighted by the local cell volume. 
As for the ensemble Kalman method, 16 samples are drawn in this case, which is the same as the square duct case. 
The convergence is achieved in about 20 iterative steps. 
In each training step, the computational cost for each sample is approximately 50 hours of CPU time (equivalent to the computational cost of the baseline or trained DD-EAS-DDES models in this case), resulting in a total training cost of 16,000 CPU hours.

\subsection{Predictions of velocity and skin friction coefficient}

\begin{figure}
    \centering
    \includegraphics[width=0.99\linewidth]{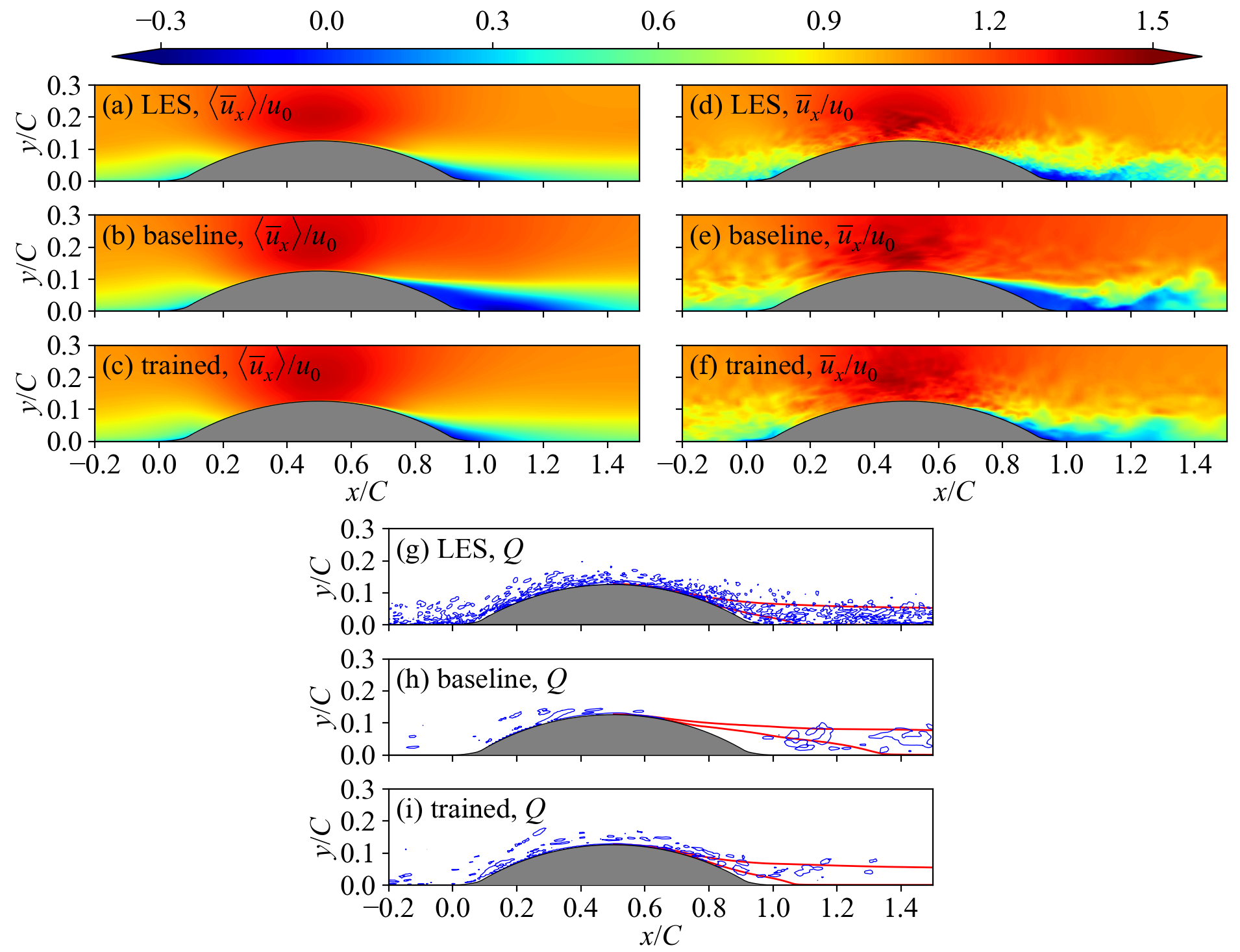}
    \caption{Comparisons of the normalized instantaneous and mean streamwise velocity contours and the Q-criteria contours at $Q = 2\times 10^5 s^{-2}$, among the (a, d) LES, (b, e) baseline EAS-DDES model, and (c, f) trained DD-EAS-DDES model for flows over the bump.}
    \label{fig:bump_contour}
\end{figure}

Figure \ref{fig:bump_contour}(a-c) shows the normalized mean streamwise velocity contours. Both the baseline and trained models are in good agreement with the LES data upstream of the bump crest, where the flow experiences a weak APG to FPG.
However, in the region of strong adverse pressure gradient after the bump crest, a substantial difference is observed in the predictions by both models.
Compared with the LES result, the baseline EAS-DDES model predicts the separation point too early, leading to a significantly large recirculation zone. 
In contrast, the prediction of the trained model is in excellent agreement with the LES result. 
Both the separation and reattachment locations are accurately predicted.

Regarding the instantaneous streamwise velocity, as shown in figure \ref{fig:bump_contour}(d-f), both the baseline and the trained models resolve the turbulent motions well before the bump crest. 
The main difference in the instantaneous velocity fields between the two models is located in the separated shear layer after the bump crest. 
The prediction of the baseline model exhibits a strong delay of transition from modelled to resolved turbulence in the separated shear layer, known as the ``grey area" issue in hybrid RANS-LES approaches \citep{probst2017evaluation, pont2021new, wang2025gray}. 
In contrast, the trained model exhibits a reasonable development of the resolved turbulent motions in the shear layer. The resolved turbulent motions occur much earlier than the baseline model and are consistent with the LES results.

The comparison is also made in the instantaneous contour lines of Q-criterion, as depicted in figure \ref{fig:bump_contour}(g-i). The blue lines denote $Q = 2\times 10^5 s^{-2}$, and the red lines mark $\langle \bar{u}_x \rangle/u_0 = 0.2, 0.8$ after the bump crest to indicate the separated shear layer. 
Both models predict larger turbulent structures than the LES due to the coarser mesh.
With the trained model, the predicted turbulent motions occur immediately upon shear layer separation, which agrees with the LES result. 
In contrast, the baseline model suffers from the ``grey area" issue, predicting a strong delay in the occurrence of turbulent motions in the separated shear layer.

\begin{figure}
    \centering
    \includegraphics[width=0.8\linewidth]{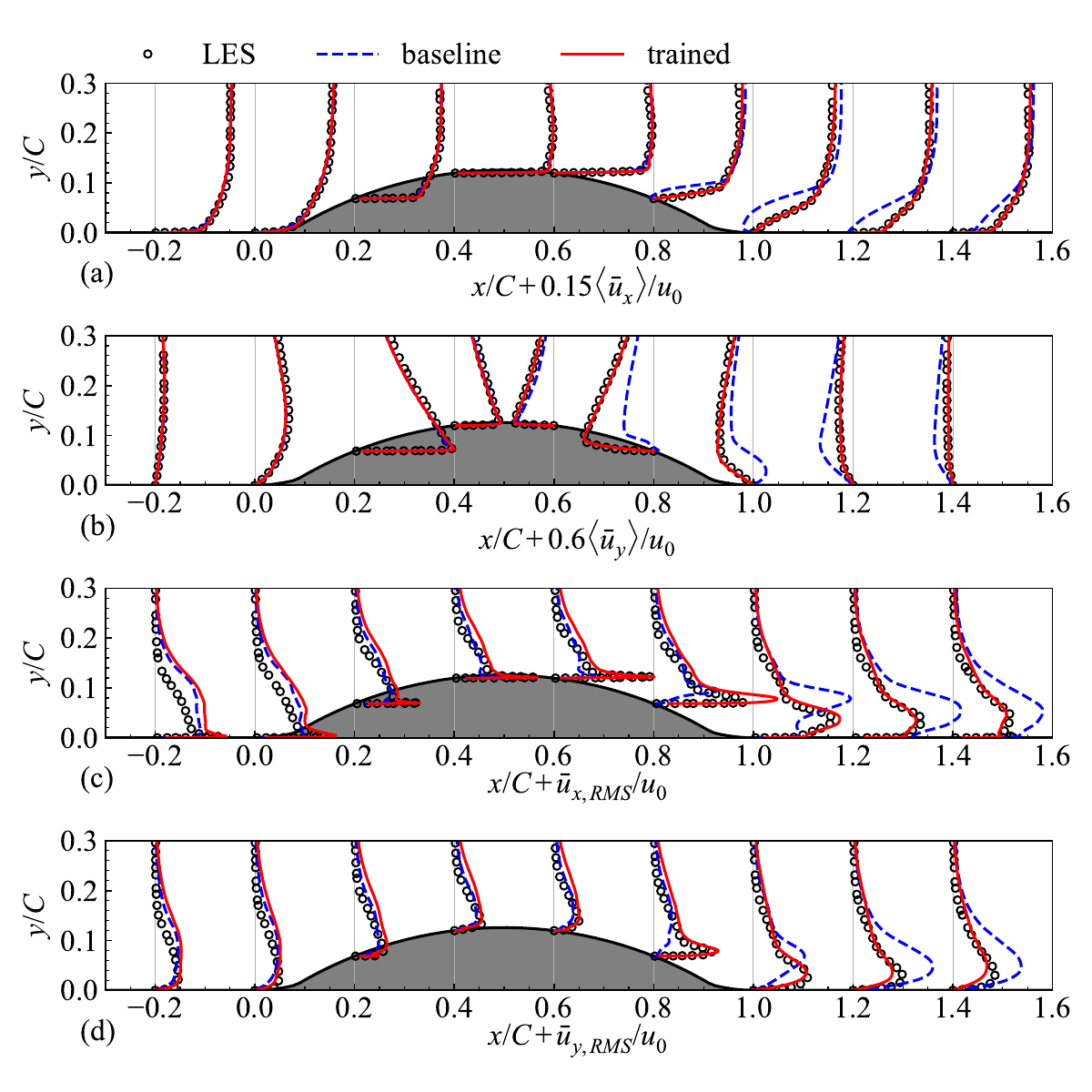}
    \caption{Comparisons of the mean and RMS velocity profiles of the bump case between the LES, the baseline EAS-DDES model, and the trained DD-EAS-DDES model at different streamwise locations with an interval of $0.2C$.}
    \label{fig:bump_U}
\end{figure}

Figure \ref{fig:bump_U}(a,b) shows the mean velocity profiles in the streamwise and wall-normal directions. 
Before the bump crest, both models can predict the velocity profiles well.
However, the baseline model exhibits large discrepancies from the LES after the bump crest due to the incorrect predictions of the separation and reattachment points. 
In contrast, the predictions of the trained model agree well with the LES results.
Although only the streamwise mean velocity is used in the loss function during the training process, the wall-normal velocity is also accurately predicted by the trained model.
This can be explained by the divergence-free constraint of the mean flow.
That is, the accurate prediction of $\langle \bar{u}_x \rangle$ inherently improves the prediction of $\langle \bar{u}_y \rangle$ in this spanwise-homogeneous case with $\bar{u}_z=0$.

The RMS velocity profiles are also significantly better predicted with the trained model than the baseline model, as shown in figure \ref{fig:bump_U}(c,d). 
At $x/C = 0.8$, the baseline model under-predicts the velocity fluctuations due to the delayed unsteadiness in the separated shear layer, as illustrated and explained in figure \ref{fig:bump_contour}. Further downstream, the velocity fluctuations are over-predicted due to the delayed reattachment of the separated shear layer.
The trained model substantially mitigates these issues.  
This significant improvement is achieved without including RMS velocities in the training loss, which further demonstrates the effectiveness of the present data-driven DES model.

\begin{figure}
    \centering
    \includegraphics[width=0.8\linewidth]{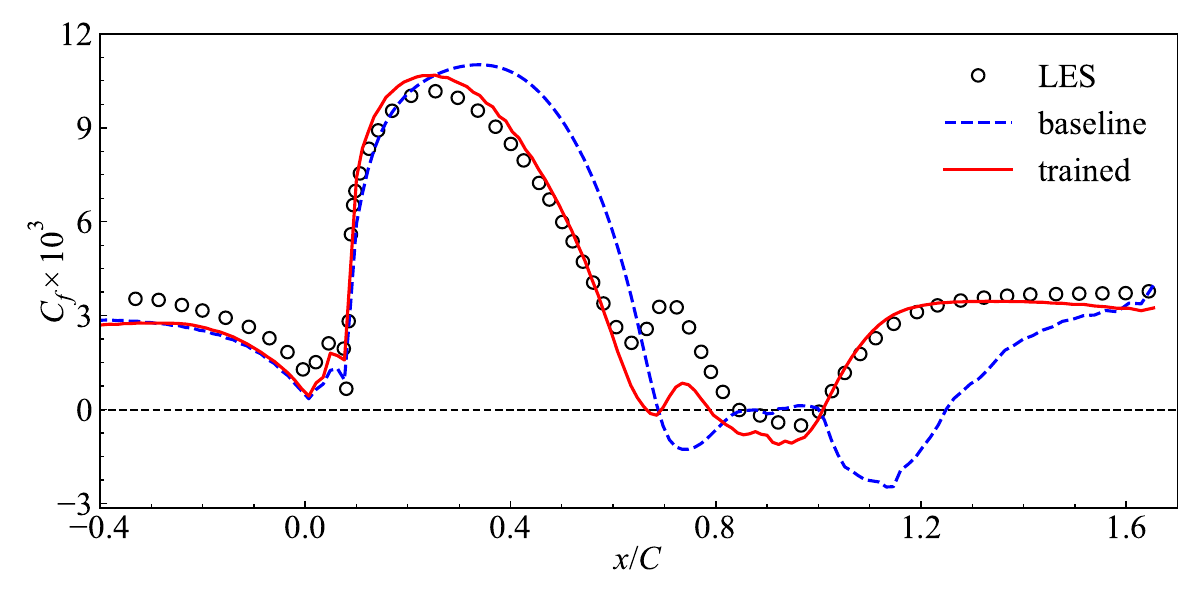}
    \caption{Comparisons of the skin friction coefficient between the LES, the baseline EAS-DDES model, and the trained DD-EAS-DDES model for flows over the bump.}
    \label{fig:bump_Cf}
\end{figure}

Figure \ref{fig:bump_Cf} illustrates the mean skin friction coefficient $C_f$ for flows over the bump.
In the region of FPG ($0 < x/C < 0.5$), the trained model accurately predicts $C_f$ in comparison with the LES result, while the baseline model over-predicts $C_f$ in the region of ($0.2 < x/C < 0.5$), and the peak location of $C_f$ is delayed in the prediction of the baseline model. 
In the APG region before the flow separation ($0.5 < x/C < 0.8$), there exists a local maximum of $C_f$ for the LES results, which is also observed in previous DNS of a Gaussian bump \citep{balin2021direct}. 
The existence of a local maximum is also captured by the trained model, although the magnitude is lower than the LES result.
In contrast, this local maximum is not observed in the prediction of the baseline model.
After the flow separation ($x/C > 0.8$), the predicted $C_f$ by the trained model agrees well with the LES results, indicating the accurate predictions of the flow separation, reattachment of the separated shear layer, and the recovery of the reattached turbulent boundary layer further downstream.
In contrast, the friction coefficient $C_f$ predicted by the baseline model severely deviates from the LES results due to the significantly over-predicted area of the separation region, as shown in figure \ref{fig:bump_contour}(d-f). 

The improvement of the predicted mean $C_f$ can be explained by the prediction of the turbulent stress. 
Based on mean friction drag decomposition in the turbulent boundary layer \citep{renard2016theoretical, fan2020decomposition, elnahhas2022enhancement}, the contributions of the mean $C_f$ consist of a laminar part, turbulent momentum flux, streamwise growth of the boundary layer, mean wall-normal flux, and the freestream pressure gradient. 
In consideration of the pressure gradient contribution, the freestream FPG and APG increase and decrease the mean $C_f$ by increasing and decreasing the freestream velocity, respectively, relative to the local boundary layer. 
This determines the variation of mean $C_f$ over the bump as shown in figure \ref{fig:bump_Cf}. 
The turbulent momentum flux contribution is determined by the turbulent shear stress in the wall frame $\hat{\tau}_{tn} \equiv \langle u_t' u_n' \rangle$, where the subscripts ``$t$'' and ``$n$'' denotes the wall-tangential and -normal directions, respectively. 
In general, the increase of $\hat{\tau}_{tn}$ leads to the increase of the mean $C_f$.

\begin{figure}
    \centering
    \includegraphics[width=0.6\linewidth]{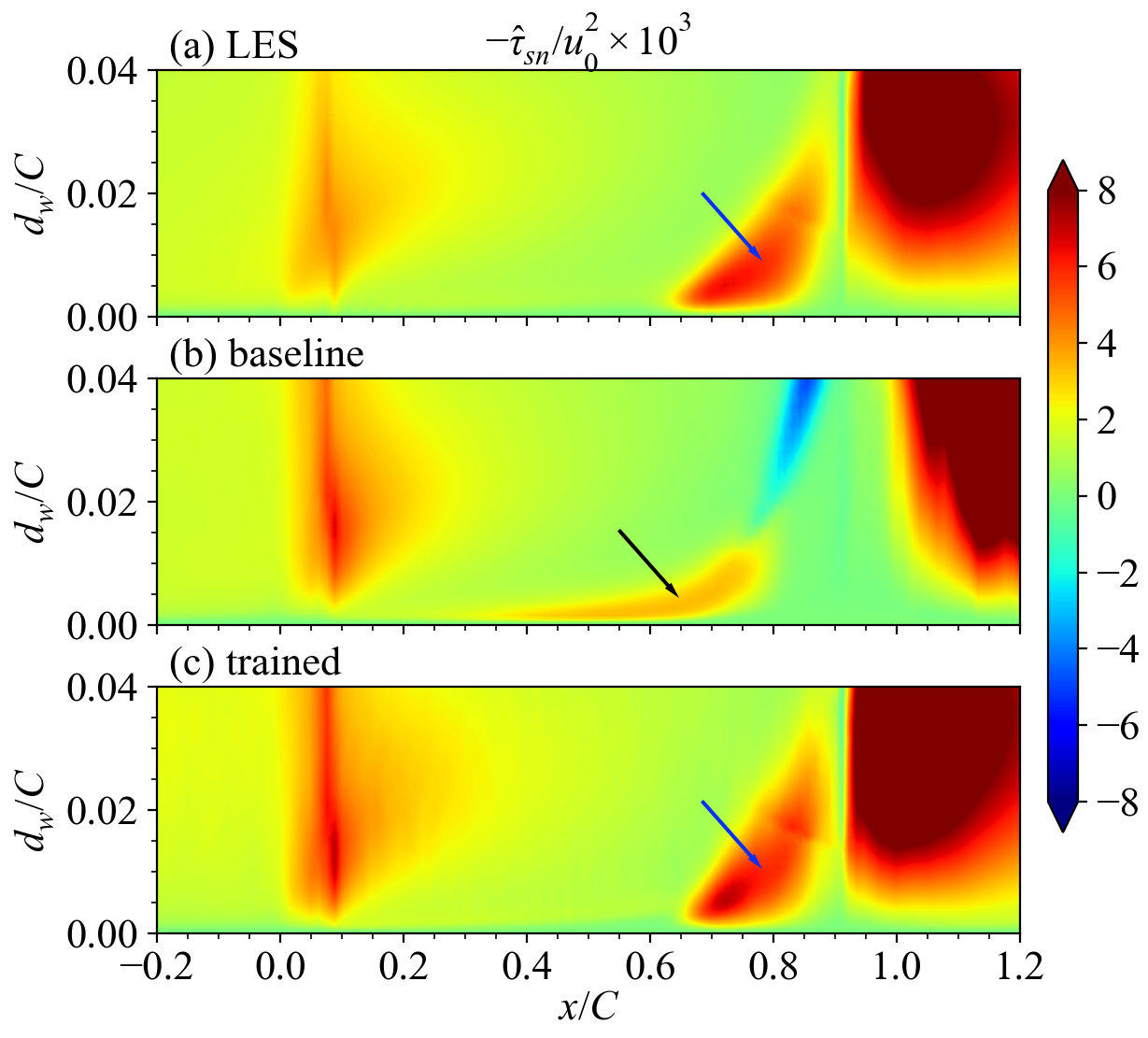}
    \caption{Comparisons of the Reynolds shear stress in the wall-frame between the LES, the baseline EAS-DDES model, and the trained DD-EAS-DDES model for flows over the bump.}
    \label{fig:bump_uv_contour}
\end{figure}

Figure \ref{fig:bump_uv_contour} depicts the distribution of $\hat{\tau}_{tn}$ under the local wall frame in the near-wall region. 
In the range of $0 < x/C < 0.2$, both models predict a slightly higher value of $\hat{\tau}_{tn}$ compared with the LES results. 
In $0.2 < x/C < 0.6$, the trained model predicts a low magnitude of $\hat{\tau}_{tn}$ near the wall due to the presence of the FPG, which is close to the LES result. 
However, the baseline model predicts a much higher $-\hat{\tau}_{tn}$ in this region (marked by the black arrow), resulting in an overestimated mean $C_f$ in this region.
Further downstream, the prediction of the trained model is similar to the LES results. 
A high turbulent shear stress region appears in the range of $0.6 < x/C < 0.9$ due to the strong APG effect, as marked by the blue arrows. 
This leads to the abrupt increase of mean $C_f$ and the occurrence of a local maximum, as shown in figure \ref{fig:bump_Cf}. 
The occurrence of this high $-\hat{\tau}_{tn}$ region is slightly delayed for the trained model compared with the LES result, leading to the slightly delayed increase of the mean $C_f$. 
The high $-\hat{\tau}_{tn}$ region enables the near-wall flows to resist the strong APG and keep attached until $x/C \approx 0.8$. 
As for the baseline model, the magnitude of $\hat{\tau}_{tn}$ in the range $0.6 < x/C < 0.8$ is significantly under-predicted, leading to premature flow separation. 
Overall, the baseline model fails to accurately reflect the effect of the pressure gradient on Reynolds shear stress, resulting in inaccurate predictions of the separation point and the mean $C_f$. 
In contrast, the trained model significantly improves the prediction of Reynolds shear stress, resulting in enhanced prediction performance.

\subsection{Switching behaviours and model coefficients}

\begin{figure}
    \centering
    \includegraphics[width=0.8\linewidth]{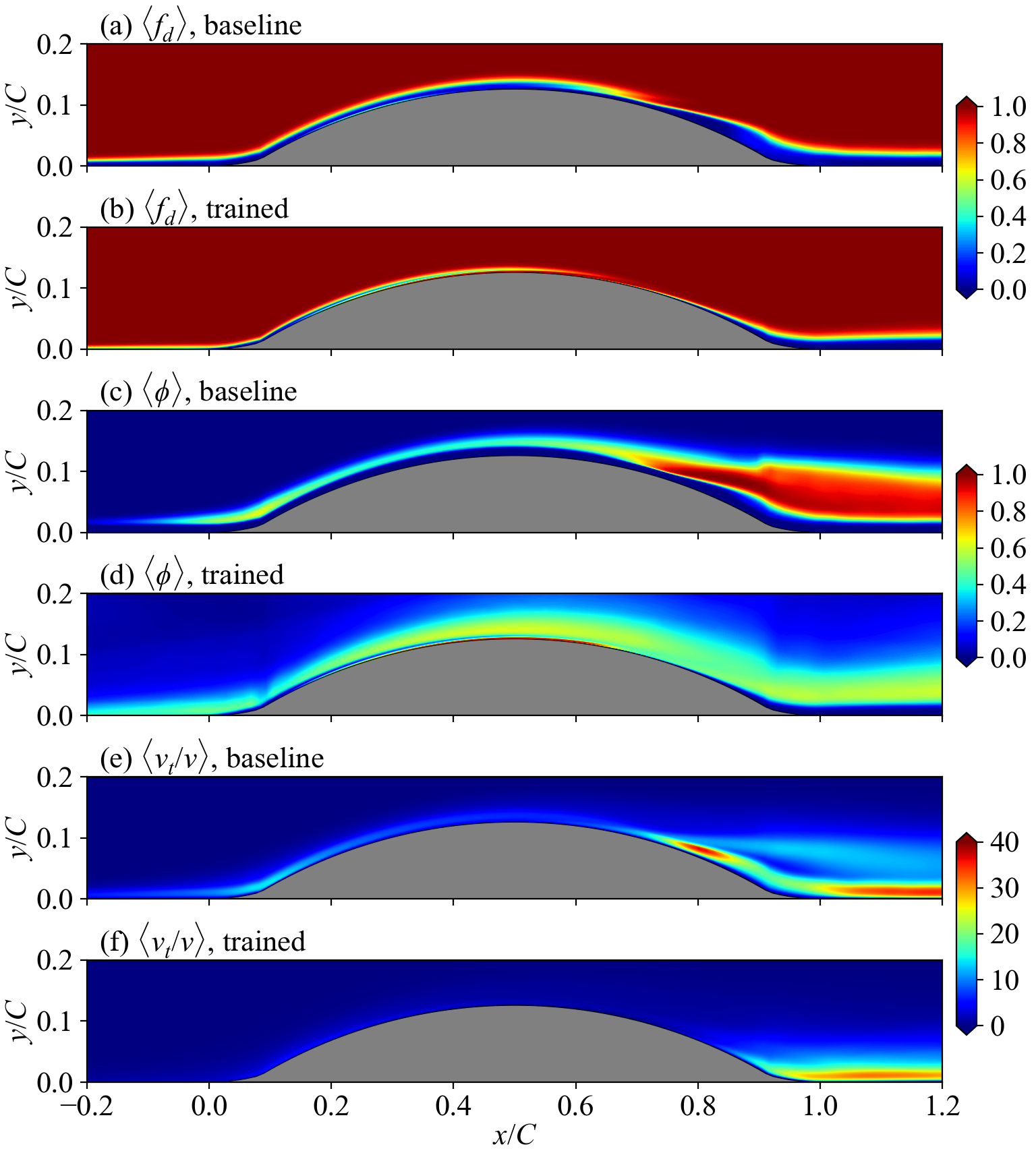}
    \caption{Comparisons of the (a, b) mean DES shielding function $\langle f_d \rangle$, (c, d) the LES mode indicator $\langle \phi \rangle$ and (e, f) the turbulent viscosity ratio $\langle \nu_t \rangle/\nu$ between the baseline EAS-DDES model and the trained DD-EAS-DDES model for flows over the bump.}
    \label{fig:bump_fd_nut}
\end{figure}

Figure \ref{fig:bump_fd_nut}(a, b) compares the mean shielding function $\langle f_d \rangle$ of the baseline and the trained models. Similar to the square duct case, the trained model reduces the near-wall shielded RANS layer in the region over the bump ($x/C<1$). 
As explained in \S \ref{sec:switching}, $f_d=1$ does not necessarily mean the model is switched to LES, and this is true for the present case due to a moderate mesh resolution ($\upDelta x^+ \approx 100$ and $\upDelta z^+ \approx 40$).
Given this, we define the LES mode indicator $\phi$ as
\begin{equation}
\label{phi}
    \phi = \frac{\ell_{DDES} - \ell_{RANS}}{\ell_{LES} - \ell_{RANS}},
\end{equation}
to further demonstrate the switching behaviour. 
This indicator~$\phi$ varies from 0 to 1, with $\phi = 0$ and $1$ corresponding to the use of RANS or LES length scale for $\ell_{DDES}$, respectively.

The mean value of $\phi$ is shown in figure \ref{fig:bump_fd_nut}(c, d). In the region before the flow separation ($x/C < 0.7$), the trained model has a higher value of $\langle \phi \rangle$ than the baseline model. 
This means that the trained model tends to resolve a larger proportion of turbulence scales to better predict the local turbulent state, yielding a significant improvement in the skin friction on the hump, as previously shown in figure \ref{fig:bump_Cf}.
More importantly, the convection of resolved turbulent structures into the near-wall boundary layer enhances its resistance against strong adverse pressure gradients (APG), resulting in delayed flow separation and a more accurate separation point prediction.
In figure \ref{fig:bump_fd_nut}(f), a clear, near-wall RANS layer is established after $x/C=1.0$, which results in a good skin friction prediction after reattachment, as shown in figure \ref{fig:bump_Cf}.

\begin{figure}
    \centering
    \includegraphics[width=0.99\linewidth]{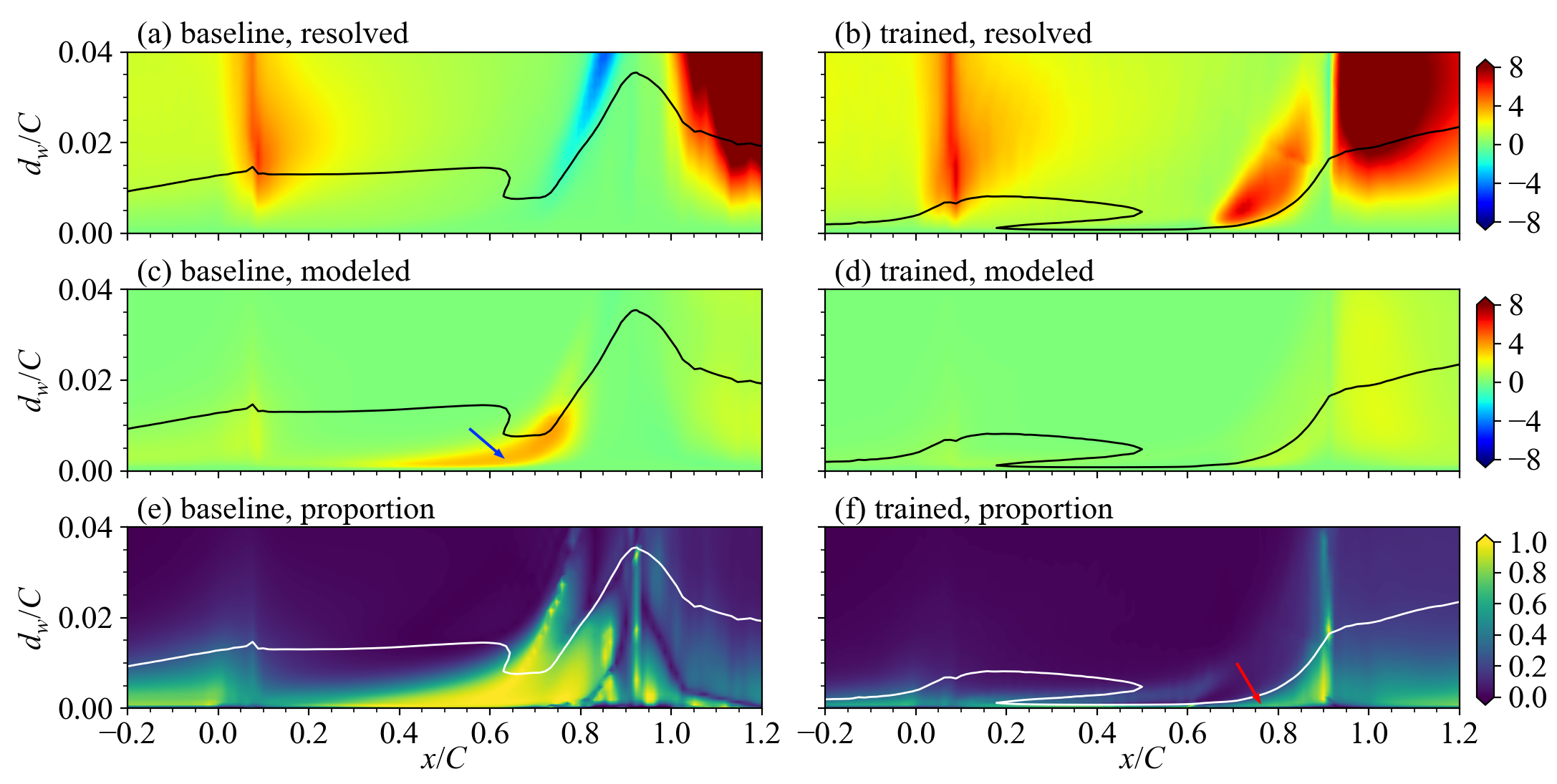}
    \caption{Comparisons of the resolved and modelled Reynolds shear stress ($\langle \bar{u}_t' \bar{u}_n' \rangle / u_0^2 \times 10^3$ and $\langle \tau_{tn} \rangle / u_0^2 \times 10^3$) in the wall-frame and the proportion of the modelled part $|\langle \tau_{tn} \rangle | /( |\langle \bar{u}_t' \bar{u}_n' \rangle| + |\langle \tau_{tn} \rangle|) $ between the LES, the baseline EAS-DDES model and the trained DD-EAS-DDES model. The solid line is the contour line of $\langle f_d \rangle = 0.5$.}
    \label{fig:bump_uv_mol_contour}
\end{figure}

Further, the resolved and modelled parts of the Reynolds shear stress are illustrated in figure \ref{fig:bump_uv_mol_contour}.
The solid line roughly indicates the RANS/LES switching location.
For the baseline model, the near-wall RANS region is relatively thick. 
Moreover, the Reynolds shear stress in $0.2 < x/C < 0.6$ is over-predicted (pointed by the blue arrow in figure \ref{fig:bump_uv_mol_contour}(c)), which can be attributed to the modelling error of the RANS branch in capturing the effects of the pressure gradient on the Reynolds stress.
For the trained model, the near-wall flow in $0.2 < x/C < 0.6$ is largely resolved, and the effects of the pressure gradient can be effectively captured. 
Further downstream, the strong APG significantly enhances the turbulent intensity, leading to an abrupt thickening of the near-wall RANS region. 
Before the separation ($0.6 < x/C < 0.8$), the proportion of the modelled stress played a crucial role in the total stress, as pointed by the red arrow in figure \ref{fig:bump_uv_mol_contour}(f). 
This indicates that the near-wall RANS model has non-negligible contributions to the improved prediction of the flow separation point in this case.

\begin{figure}
    \centering
    \includegraphics[width=0.8\linewidth]{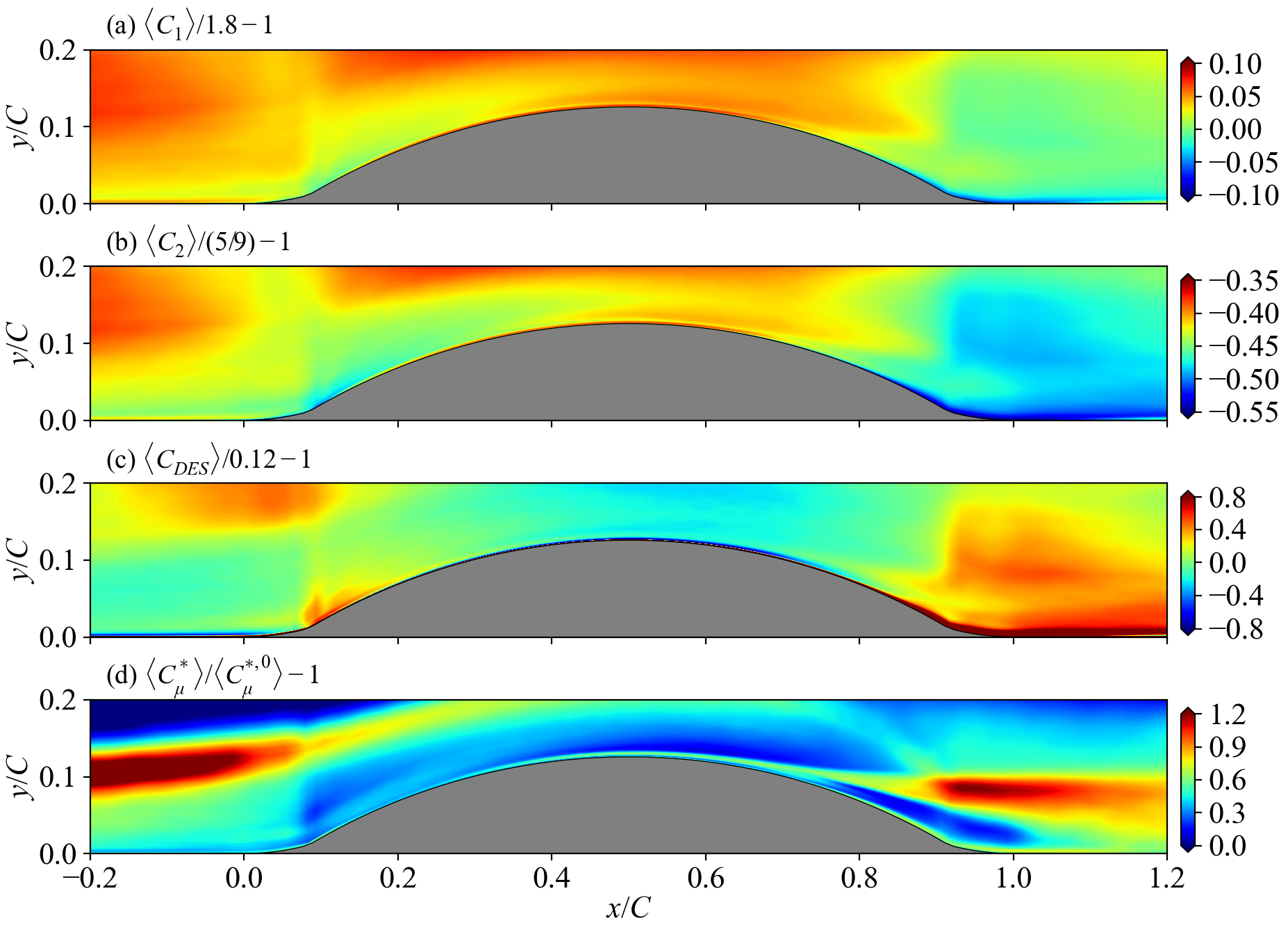}
    \caption{Mean values of the model coefficients $\langle C_1 \rangle$, $\langle C_2 \rangle$, $\langle C_\mathrm{DES} \rangle$, and $\langle  C_\mu^* \rangle$ of the trained DD-EAS-DDES model.}
    \label{fig:bump_C_contour}
\end{figure}

The trained model coefficients $C_1$, $C_2$, and $C_\mathrm{DES}$, and the calculated $C_\mu^*$ are examined for this case to explain the model behaviour. 
The relative variations of the trained coefficients compared to the baseline values are shown in figure \ref{fig:bump_C_contour}. 
The value of $\langle C_\mathrm{DES} \rangle$ is lower than the baseline value in the near-wall region before the separation ($x/C < 0.7$). 
A smaller $\langle C_\mathrm{DES} \rangle$ predicted by the trained model leads to decreased subgrid viscosity (plotted in figure \ref{fig:bump_fd_nut}), and thus reduces the production of $k$ near the edge of the RANS region.
The variation of $k$ can affect the RANS region thickness due to the dependency of $f_d$ on $k$ as in equation~\eqref{fd}.
This is the main reason for the reduction of the near-wall shielded RANS region. 
Additionally, the mean $C_\mu^*$ is larger than the baseline value in the whole domain through the variations in $C_1$ and $C_2$ from their baseline values. 
This modification to the near-wall RANS model can play an important role in the strong APG region, due to the abrupt thickening of the near-wall RANS region and the increase in the modelled stress contribution, as demonstrated in figure \ref{fig:bump_uv_mol_contour}. 
The relatively large $C_\mu^*$ increases the near-wall shear stress in this region, enabling the flow to resist the APG and keep attached until the correct separation point. 
To summarize, the combined modifications of $C_\mathrm{DES}$ and $C_\mu^*$ (modified through $C_1$ and $C_2$) lead to the effective response of the near-wall shear stress to the pressure gradient by enlarging the LES region in the FPG region and modifying the RANS model in the APG region, which improves the model prediction of the flow separation point and the skin friction.

\subsection{Generalization in similar geometries and different grid resolutions}

The generalizability of the trained model is further examined by simulating flows over the bumps with different heights. 
Two different bump heights, i.e., $h = 26$ and 42 mm, are chosen, denoted by the h26 and h42 cases, respectively. 
The h26 case is marked as the ``flow on verge of separation" in the LES of \citet{matai2019large}. 
The flows over the two bumps are simulated using the baseline and trained models, respectively, with comparison to the LES results. 
The employed trained model is trained from the h38 case with the bump height $h=38$ mm. 
The mesh quality is kept the same as the training case.

\begin{figure}
    \centering
    \includegraphics[width=0.99\linewidth]{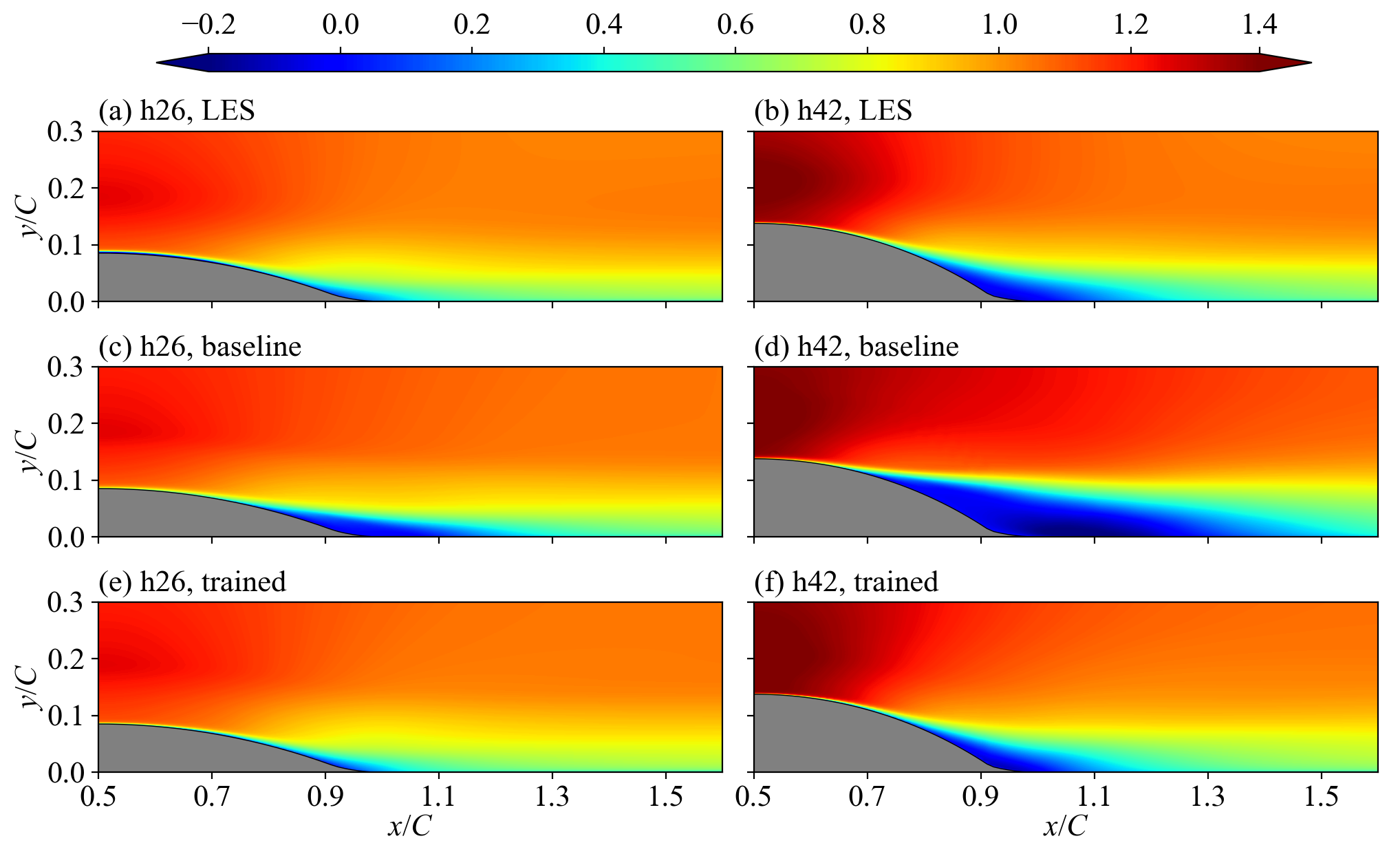}
    \caption{Comparisons of the normalized mean streamwise velocity $\langle \bar{u}_x \rangle / u_{0}$ contours of the flow over the bump between (a, b) the LES, (c, d) the baseline EAS-DDES model, and (e, f) the trained DD-EAS-DDES model for the h26 and the h42 bumps. 
    The DD-EAS-DDES model is trained in the h38 bump case.}
    \label{fig:bump_general_U}
\end{figure}

Figure \ref{fig:bump_general_U} illustrates the normalized mean streamwise velocity $\langle \bar{u}_x \rangle / u_{0}$. 
It can be observed that the baseline model predicts a premature separation and a delayed flow reattachment for both cases, which is similar to the observed features in the h38 case. 
The trained model predicts very similar flow patterns as the LES results for both cases.
The separation and reattachment points of the h42 case are correctly predicted. 
The flow features on the verge of separation for the h26 case are also accurately captured with the trained model.

\begin{figure}
    \centering
    \includegraphics[width=0.8\linewidth]{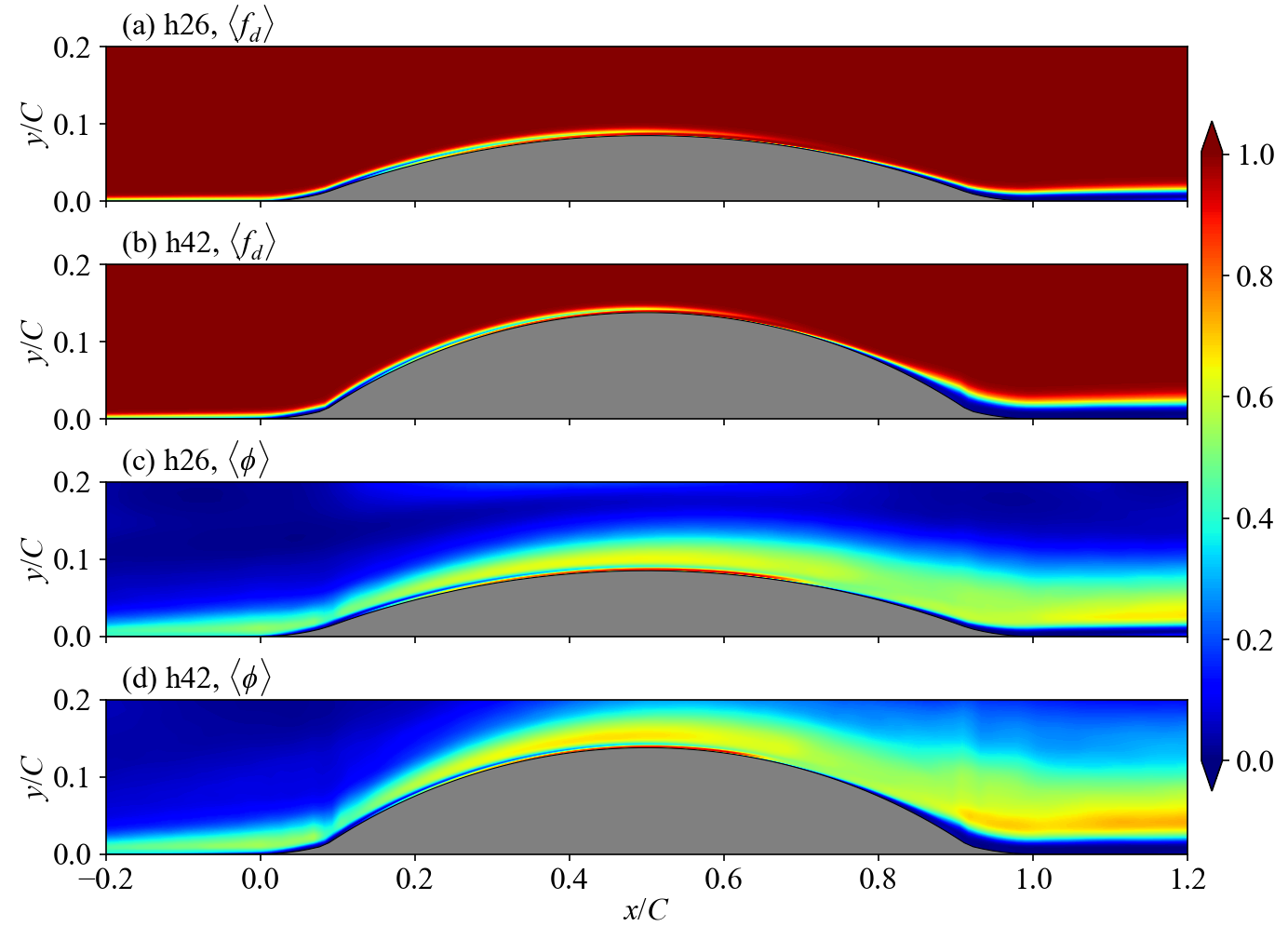}
    \caption{The mean shielding function $\langle f_d \rangle$ and LES mode indicator $\langle \phi \rangle$ contours simulated by the trained DD-EAS-DDES model for the h26 and the h42 bumps. 
    The DD-EAS-DDES model is trained in the h38 bump case.}
    \label{fig:bump_fd_contour_generalize}
\end{figure}

Figure \ref{fig:bump_fd_contour_generalize} presents the predicted mean shielding function $\langle f_d \rangle$ and LES mode indicator $\langle \phi \rangle$ contours for the h26 and h42 bumps. These results closely match the performance observed in the training case (the h38 bump, shown in figure \ref{fig:bump_fd_nut}(b,d)). Therefore, the switching behaviour of the trained DD-EAS-DDES model can be generalized to geometrically similar configurations.

\begin{figure}
    \centering
    \includegraphics[width=0.8\linewidth]{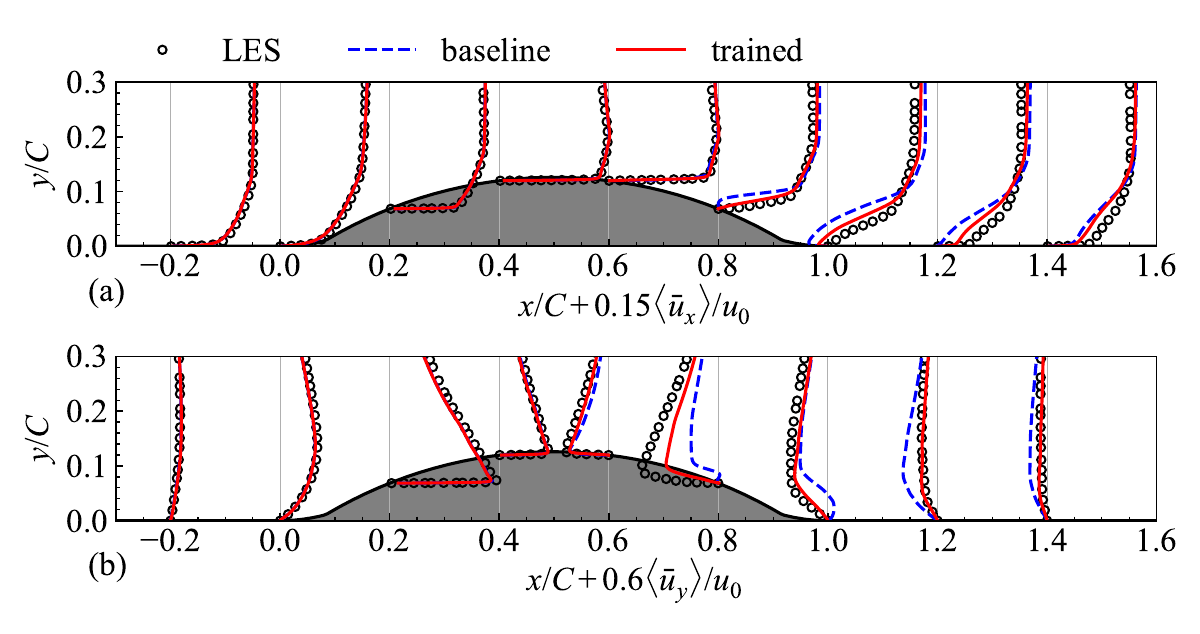}
    \caption{Comparisons of the mean velocity profiles of the bump case between the LES, the baseline EAS-DDES model, and the trained DD-EAS-DDES model at different streamwise locations with an interval of $0.2C$. The DES cases are under the fine mesh, while the trained model is trained with the base mesh.}
    \label{fig:bump_U_general_mesh}
\end{figure}

In addition, the trained model's generalizability has been assessed on a finer mesh, with the cell sizes halved in both the streamwise and spanwise directions, corresponding to $\upDelta x^+ \approx 50$ and $\upDelta z^+ \approx 20$ in the bump region.
The cell size in the wall-normal direction is maintained.
The DES is performed with the baseline and the trained models, where the latter is trained on the base mesh. 
The mean velocity profiles are depicted in figure \ref{fig:bump_U_general_mesh}. 
Similar to the base-mesh results, the trained model performs significantly better compared to the baseline model on the fine mesh, indicating the robustness of the trained DD-EAS-DDES approach under different mesh resolutions. 
It is also worth noting that for the trained model, the fine-mesh results are relatively less accurate than those on the base mesh, as shown in figure \ref{fig:bump_U}(a,b). 
This could be further improved by training the model with different mesh resolutions.

\begin{figure}
    \centering
    \includegraphics[width=0.8\linewidth]{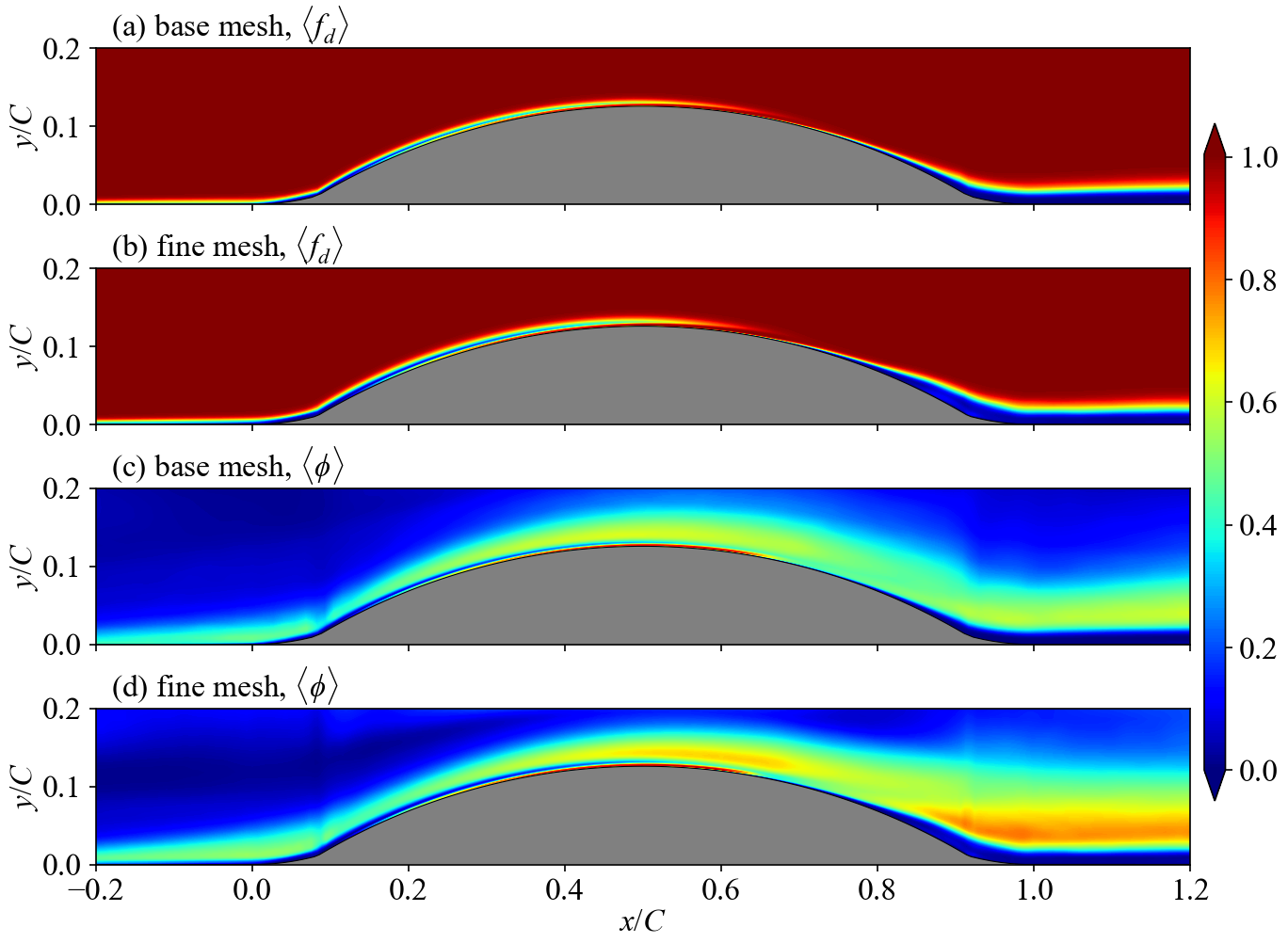}
    \caption{The mean shielding function $\langle f_d \rangle$ and LES mode indicator $\langle \phi \rangle$ contours simulated by the trained DD-EAS-DDES model for different mesh resolutions. 
    The DD-EAS-DDES model is trained with the base mesh.}
    \label{fig:bump_fd_contour_generalize_mesh}
\end{figure}

The switching behaviours of the trained model under different mesh resolutions are demonstrated in figure \ref{fig:bump_fd_contour_generalize_mesh}. The mesh resolution has negligible influence on the DDES shielding function $f_d$. 
It is reasonable because the shielded RANS region is already very thin, especially on the bump surface.
Since switching too close to the wall would cause premature separation, the trained model seems to sustain a robust shielding for the RANS region.
In contrast to $f_d$, the LES mode indicator $\langle \phi \rangle$ reflects the local proportion of the resolved turbulent scales. 
With finer meshes, the LES mode is more prone to be activated outside the near-wall shielded region due to the decrease of $\ell_{LES}$. 
This trend is well demonstrated by the plot of $\langle \phi \rangle$ in figure \ref{fig:bump_fd_contour_generalize_mesh}(c,d). 
Therefore, the trained model has a physically justified switching behaviour in the DDES framework.

\section{Discussion}
\label{sec:discussion}

\subsection{Physical realizability}

The present model is formulated based on the algebraic equation of the modelled stress anisotropy, which ensures that the obtained modelled stress satisfies the weak-equilibrium assumption.
This is found to be a reasonable approximation of the full differential Reynolds stress transport equations in various flow scenarios \citep{wallin2000explicit}. 
Hence, in the DD-EAS-DDES model, the obtained modelled stress incorporates the physical constraint of the algebraic stress equation, which captures the physical process of the local turbulence under the weak-equilibrium assumption.
In contrast, conventional neural network-based models do not embed such physical constraints, such as the TBNN model, where the neural network represents the mapping from the flow features to the tensor basis coefficients $\beta_i$.

One advantage of the inherent physical constraint in the DD-EAS-DDES model is the satisfaction of physical realizability.
The EARS model \citep{wallin2000explicit} is primarily formulated from the algebraic Reynolds stress equation using the LRR model~\citep{launder1975progress} for the pressure-strain rate terms. 
The LRR model is known to satisfy weak realizability in all but pathologic circumstances \citep{durbin1994realizability, pope2000turbulent}, and the current model inherits this desirable property. 
Regarding the trained model, the regularization applied to the coefficients $C_1$, $C_2$ and $C_{DES}$, as described in \S \ref{sec:training}, ensures that these values do not deviate significantly from those of the baseline model. 
This also helps the trained model to preserve the realizability from the baseline model.

\begin{figure}
    \centering
    \includegraphics[width=0.8\linewidth]{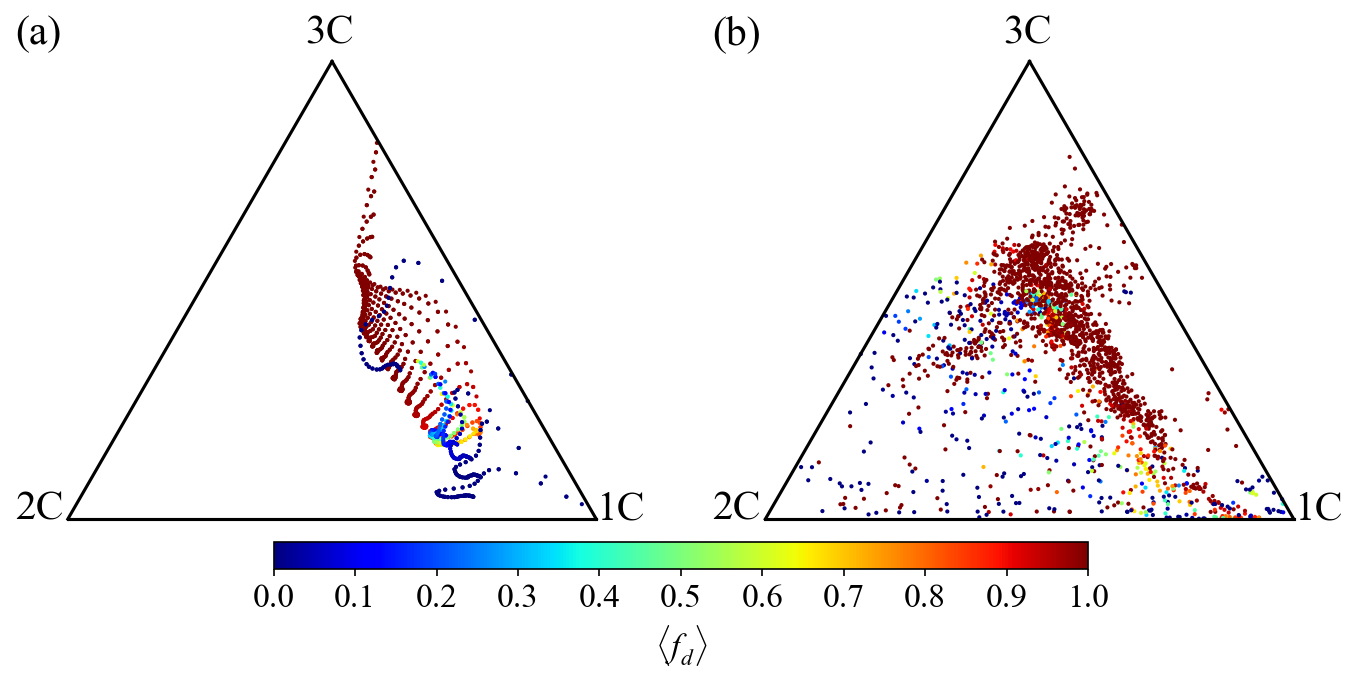}
    \caption{The local state of the Reynolds stress anisotropy in the Barycentric triangle for (a) the square duct flow and (b) the flow over the bump. The points are coloured by the mean DES shielding function $\langle f_d \rangle$.}
    \label{fig:Barycentric}
\end{figure}

The realizability behaviour is demonstrated in Figure \ref{fig:Barycentric} for the square duct and the bump cases simulated by the trained model.
The data points on a homogeneous plane for both cases are plotted in Barycentric coordinates coloured by the mean shielding function $\langle f_d \rangle$, where the Barycentric triangle encloses all physically realizable states \citep{banerjee2007presentation}. For the bump case, only one-fifth of the points are plotted for clarity.
It can be seen that all the points fall inside the Barycentric triangle. 
The physical realizability of the simulated state of turbulent stress by the trained DD-EAS-DDES model in both LES and RANS branches is verified.
For other modelling approaches, extra regularization has to be introduced during the training process to ensure physical realizability~\citep{jiang2021interpretable}.

\subsection{Training efficiency}

Using the EAS-based turbulence model as the baseline can facilitate training efficiency with improved accuracy.
The LEV model is typically considered the baseline or initialized model before training for many data-driven turbulence models. 
For instance, in the TBNN model, the initial weights are typically set to ensure $C_\mu^* = 0.09$ and $\beta_i = 0 \quad (i = 2, \dots, 10)$ \citep{zhang_ensemble-based_2022}. 
This corresponds to the conventional LEV-RANS models, e.g., standard $k-\varepsilon$ or $k-\omega$ models, or a corresponding LEV-DES model if the TBNN framework is adopted in formulating a data-driven DES model.
The non-linear component of the modelled stress is ignored in the initialized model and has to be developed from scratch with the data-driven method. 
In contrast, the present approach uses the baseline EAS-DDES model to initialize the neural network weights.
The superiority of this model in predictive accuracy compared to the LEV-DES models has been demonstrated \citep{liu2024explicit}. 
The non-linear component of the modelled stress is inherently accounted for and can be further modified efficiently to achieve better accuracy through the modification of $C_1, C_2$, and $C_\mathrm{DES}$ during the training process.
Therefore, the present algebraic stress-based approach can produce a trained model with better efficiency than models initialized as LEV models.

Another obvious advantage of the present approach is that the number of neural network weights required to be trained is relatively small.
Although the modelled stress includes 10 tensor bases, only 3 model coefficients are required to be trained, and the 10 tensor basis coefficients $\beta_i$ can be obtained by solving the algebraic equation of the modelled stress anisotropy.
This is in contrast to the TBNN model, where the 10 tensor basis coefficients are required to be trained for the most general form of the modelled stress anisotropy.
In the RANS framework, the tensor basis coefficients can be reduced to only two for statistically two-dimensional flows \citep{pope2000turbulent}, while this simplicity cannot be inherited in the eddy-resolved simulations, e.g., DES for the present study, due to the three-dimensional nature of turbulence. 
Hence, the present approach can reduce the neural network weights to achieve improved training efficiency.

\subsection{Computational efficiency}

The present DD-EAS-DDES model computes the Reynolds stress anisotropy from algebraic formulations and uses a neural network to represent the model coefficients,
without introducing extra transport equations. 
Hence, the present model maintains the computational efficiency of the baseline EAS-DDES model, or any existing $k-\omega$ based DES approaches, which is significantly less costly than the WRLES or DNS.
Specifically, for the present simulation of the flow over a bump, the mesh resolution is $\upDelta x^+ \approx 100$ and $\upDelta z^+ \approx 40$ in the bump region, which is approximately 3 times coarser per direction than the WRLES \citep{matai2019large}, and 6 times coarser in wall units than the DNS in a similar case \citep{balin2021direct}. 
Considering the related requirement of the time step size, the present DD-EAS-DDES reduces the computational cost by factors of 30 and 300 compared with the corresponding WRLES and DNS, respectively. 
Furthermore, the near-wall RANS behaviour for DES largely relaxes the wall-normal cell spacing requirement compared with LES and DNS, leading to further reduction of the computational cost. 

As another illustration, \citet{spalart2000strategies} provided a rough estimation of the computational cost for different turbulence simulation strategies. The targets are high-Reynolds-number external flows typical of commercial aircraft and ground vehicles. The evaluated computational costs for DES, LES, and DNS, in terms of spatial and temporal degrees of freedom, are $10^{12}$, $10^{18.2}$, and $10^{23.7}$, respectively. 
This highlights the advantage of DES in computational efficiency over LES or DNS for practical applications.

\section{Conclusion}
\label{sec:conclusion}

In the present work, a data-driven DES method based on the algebraic stress equation is proposed. 
The modelled stress in the DES is formulated as the summation of the linear and non-linear parts. 
The linear part is responsible for the switching between the RANS and LES branches through the eddy viscosity $\nu_t$, based on the $\ell^2-\omega$ framework. 
The non-linear part introduces additional anisotropy to the modelled stress. 
The modelled stress anisotropy is expressed as the tensor basis form, with the tensor basis coefficients $\beta_i$ obtained by solving the algebraic equation of the modelled stress anisotropy under the weak-equilibrium assumptions. 
There exist three model coefficients, i.e., $C_1$ and $C_2$ for the slow and rapid pressure-strain rate terms in the algebraic stress equation, and $C_\mathrm{DES}$ in the LES length scale.
These coefficients are formulated as functions of the scalar invariants $\theta_i$, represented with the neural network.
The neural network is trained by the ensemble Kalman method with velocity data, circumventing the need to provide the modelled stress data.
Moreover, to ensure a similar RANS/LES switching behaviour as the baseline model, the model coefficients $C_1$, $C_2$, and $C_\mathrm{DES}$ are augmented to the observation space, with the baseline values treated as the reference data.

The capability of the proposed framework is illustrated in two challenging turbulent flows: the secondary flow in a square duct and the separated flow over a bump. 
The former is characterized by significant Reynolds stress anisotropies that induce secondary motions, and the latter is featured by strong pressure gradients in the turbulent boundary layer and the flow separation on a curved surface.
The DNS for the square duct case and the WRLES for the bump case are used as the training data. 
For both cases, the trained model achieves significant improvements compared with the baseline model in predicting the mean flow statistics, including mean velocity and skin friction coefficient.
Such improvements are mainly attributed to the improved modelled stress.
Moreover, reasonable switching behaviour between the RANS and LES modes is obtained, where the trained model tends to enlarge the LES region to resolve more turbulent structures. 
The distributions of the trained model coefficients $C_1$, $C_2$, and $C_\mathrm{DES}$ are analyzed to explain the model behaviour. The combined effects of the adaptive $C_\mathrm{DES}$ and $C_\mu^*$ (controlled by $C_1$ and $C_2$) contribute to the expansion of the LES region. It is also observed that the adaptive coefficients improve the modelled stress predictions on both the RANS and LES branches.
The generalizability of the trained model is further examined by simulating the square duct flow at a relatively high Reynolds number and the bump flow with different crest heights. 
The predictions of the trained model in all these cases are in better agreement with the DNS or WRLES results, compared to the baseline model.

A main limitation of this study is that two distinct turbulence models (with the same structure but different neural network weight parameters $\boldsymbol{w}$) are trained separately for two different flow cases, rather than developing a single, universal turbulence model applicable to a broader range of scenarios. 
We note that the primary objective of this work is to establish a data-driven framework for DES based on the algebraic stress equation. 
Therefore, the development of a unified model across multiple cases is not pursued here. 
This work serves as a foundational step toward integrating data-driven methods with DES, and extending the framework to handle multiple flow configurations will be the focus of future investigation.
Future work will also focus on generalizing the data-driven DES model to significantly different Reynolds numbers and geometries from the training data. Special attention will be paid to the model's switching behaviour during the generalization, and the consistency and convergence behaviours of the RANS-to-LES interface will be rigorously examined.

Another potential concern regarding the current DD-EAS-DDES approach is its relatively high training cost. This work primarily focuses on establishing a novel data-driven DES framework. While reducing the training expense of a three-dimensional, unsteady data-driven turbulence model remains an important goal for future research, it is not the focus of this work. 
Notably, the trained model demonstrates satisfactory generalizability to flows with similar geometries and Reynolds numbers. This capability allows the initial training cost to be amortized across numerous simulations, e.g., in parametric studies or design optimization, thereby enhancing overall computational efficiency over time. 
Therefore, the approach can be highly cost-effective for long-term use.

\begin{bmhead}[Acknowledgements.]
This work is supported by NSFC the Excellence Research Group Program for multiscale problems in nonlinear mechanics (Grant No. 12588201).
Hao-Chen Liu also acknowledges support from the China National Postdoctoral Program for Innovative Talents (No.~BX20250282) and the China Postdoctoral Science Foundation
(No.~2024M763359). 
Xin-Lei Zhang acknowledges support from the Young Elite Scientists Sponsorship Program by CAST (No. 2022QNRC001) and the CAS Project for Young Scientists in Basic Research (YSBR-087). Finally, the authors gratefully acknowledge Dr. Racheet Matai for providing the LES data of the bump case.
\end{bmhead}

\begin{bmhead}[Declaration of interests.]
The authors report no conflict of interest.
\end{bmhead}

\begin{appen}

\section{Full expressions of the tensor basis coefficients}\label{sec:appA}

The solution of the algebraic equation \eqref{ARSM1} of the modelled stress anisotropy is provided here. 
The coefficients $A_i$ are defined for convenience, which are related to the pressure-strain rate coefficients $C_1$ and $C_2$ as
\begin{equation}
    A_1 = \frac{88}{15(7C_2+1)}, \quad 
    A_2 = \frac{5-9C_2}{7C_2+1}, \quad
    A_3 = \frac{11(C_1 - 1)}{7 C_2 + 1}, \quad
    A_4 = \frac{11}{7C_2+1} .
\end{equation}
The parameter $N$ is defined as a function of $\mathcal{P}/\varepsilon$,
\begin{equation}
\label{N}
    N = A_3 + A_4 \frac{\mathcal{P}}{\varepsilon} .
\end{equation}

Inserting equation \eqref{afull} into \eqref{ARSM1}, the coefficients $\beta_i$ can be obtained as a function of $N$ as
\begin{equation}
\begin{aligned}
& \beta_1=-\frac{1}{2} A_1 N\left(30 A_2 \theta_4-21 N \theta_2-2 A_2^3 \theta_3+6 N^3-3 A_2^2 \theta_1 N\right) / Q, \\
& \beta_2=-A_1 A_2\left(6 A_2 \theta_4+12 N \theta_2+2 A_2^3 \theta_3-6 N^3+3 A_2^2 \theta_1 N\right) / Q, \\
& \beta_3=-3 A_1\left(2 A_2^2 \theta_3+3 N A_2 \theta_1+6 \theta_4\right) / Q, \\
& \beta_4=-A_1\left(2 A_2^3 \theta_3+3 A_2^2 N \theta_1+6 A_2 \theta_4-6 N \theta_2+3 N^3\right) / Q, \\
& \beta_5=9 A_1 A_2 N^2 / Q, \quad \beta_6=-9 A_1 N^2 / Q, \quad \beta_7=18 A_1 A_2 N / Q, \\
& \beta_8=9 A_1 A_2^2 N / Q, \quad \beta_9=9 A_1 N / Q, \quad \beta_{10}=0,
\end{aligned}
\end{equation}
with the denominator
\begin{equation}
\begin{aligned}
Q & =  3 N^5+\left(-\tfrac{15}{2} \theta_2-\tfrac{7}{2} A_2^2 \theta_1\right) N^3+\left(21 A_2 \theta_4-A_2^3 \theta_3\right) N^2 \\
& +\left(3 \theta_2^2-8  A_2^2 \theta_1 \theta_2 +24 A_2^2 \theta_5 +A_2^4 \theta_1^2\right) N+\tfrac{2}{3} A_2^5 \theta_1 \theta_3 \\
&+2 A_2^3 \theta_4 \theta_1-2 A_2^3 \theta_2 \theta_3-6 A_2 \theta_4  \theta_2 .
\end{aligned}
\end{equation}

The parameter $N$ related to the production-dissipation ratio as equation \eqref{N} can be computed based on the definition of $\mathcal{P}/\varepsilon$ as equation \eqref{pe}.
This results in a cubic equation for the two-dimensional flow and a sixth-order equation for the three-dimensional flow. 
As in the previous study \citep{liu2024explicit}, the two-dimensional solution, as given by equation \eqref{Nsolution}, is used as an approximation for the three-dimensional flow here. 
This has been claimed to induce negligible differences to the simulation results \citep{wallin2000explicit}.
\begin{equation}
\label{Nsolution}
N= \begin{cases}\frac{A_3}{3}+\left(P_1+\sqrt{P_2}\right)^{1 / 3}+\operatorname{sign}\left(P_1-\sqrt{P_2}\right)\left|P_1-\sqrt{P_2}\right|^{1 / 3}, & P_2 \geqslant 0 \\ \frac{A_3}{3}+2\left(P_1^2-P_2\right)^{1 / 6} \cos \left(\frac{1}{3} \arccos \left(\frac{P_1}{\sqrt{P_1^2-P_2}}\right)\right), & P_2<0\end{cases}
\end{equation}
where 
\begin{equation}
\begin{gathered}
P_1=\left[\frac{A_3^2}{27}+\left(\frac{A_1 A_4}{6}-\frac{2}{9} A_2^2\right) \theta_1-\frac{2}{3} \theta_2 \right] A_3, \\
P_2=P_1^2-\left[\frac{A_3^2}{9}+\left(\frac{A_1 A_4}{3}+\frac{2}{9} A_2^2\right) \theta_1+\frac{2}{3} \theta_2 \right]^3.
\end{gathered}
\end{equation}

The algebraic equation of $\boldsymbol{a}$ requires correction to obtain a reasonable value of $C_\mu^*$ at small $\mathcal{P}/\varepsilon$, where the advection and diffusion of $\boldsymbol{a}$ are non-negligible. 
In analogy to the approach of \citep{wallin2000explicit}, an effective diffusion term $\boldsymbol{D}^*$ is added to the right hand side of equation \eqref{ARSM} as the correction, which reads
\begin{equation}
    \boldsymbol{D}^* = C_D \boldsymbol{a} \bnabla \cdot \boldsymbol{T}_k ,
\end{equation}
where $\boldsymbol{T}_k$ is the flux of TKE. 
Considering the balance of the $k$-equation, we have
\begin{equation}
    \bnabla \cdot \boldsymbol{T}_k \approx \mathcal{P} - \varepsilon .
\end{equation}
Insert it into equation \eqref{ARSM1}, estimate $\mathcal{P}/\varepsilon$ by $-\beta^{eq}_1 \theta_1$ and switch off the correction at $\mathcal{P}/\varepsilon > 1$. 
Afterward, the correction of $A_3$ is obtained as
\begin{equation}
    A_3' = A_3 + \frac{11}{7 C_2 + 1} C_D  \max\left( 1+\beta^{eq}_1 \theta_1, 0 \right) 
\end{equation}
where $\beta^{eq}_1$ is the value of $\beta$ at $\mathcal{P} = \varepsilon$. 
It is estimated by the expression in a two-dimensional condition as
\begin{equation}
    \beta^{eq}_1 = - \frac{A_1 N^{eq}}{(N^{eq})^2 - 2\theta_2 - (2/3) A_2^2 \theta_1} ,
\end{equation}
where $N^{eq} = A_3 + A_4$ is the value of $N$ at $\mathcal{P} = \varepsilon$.
As for the effective diffusion coefficient $C_D$, it is determined by the condition of $C_\mu^* = C_\mu$ at zero
strain rates. 
The final expression is 
\begin{equation}
    C_D = \frac{4}{15C_\mu} - C_1 + 1
\end{equation}

\end{appen}

\bibliographystyle{jfm}
\bibliography{main}

%\nolinenumbers 
\end{document}